\documentclass[twocolumn,]{aastex63}
\usepackage[T1]{fontenc}

\usepackage{amsmath}	
\usepackage{amssymb}	
\usepackage[titletoc]{appendix}
\usepackage{import}
\usepackage[flushleft]{threeparttable}
\usepackage{array,booktabs,makecell,cellspace}
\usepackage{enumitem,kantlipsum}

\def \s {\text{s}}

\def \sh {\text{sh}}

\def \cold {\text{cold}}
\def \f {\text{f}}
\def \h {\text{h}}
\def \p {\text{p}}
\def \n {\text{n}}
\def \s {\text{s}}
\def \b {\text{b}}
\def \h {\text{h}}
\def \i {\text{i}}
\def \d {\text{d}}
\def \c {\text{c}}

\def \ej {\text{ej}}

\def \ad {\text{ad}}
\def \bo {\text{bo}}
\def \DC {\text{DC}}
\def \ff {\text{ff}}
\def \A {\text{A}}

\def \WE {\text{WE}}

\def \e {\text{e}}
\def \tot {\text{tot}}

\def \e {\text{e}}
\def \W {\text{W}}
\def \F {\text{F}}
\def \m {\text{m}}
\def \T {\text{T}}
\def \t {\text{t}}

\def \sh {\text{sh}}

\def \tr {\text{tr}}
\def \obs {\text{obs}}
\def \dyn {\text{dyn}}

\def \B {\text{B}}
\def \NW {\text{NW}}
\def \pl {\text{pl}}
\def \b {\text{b}}
\def \KN {\text{KN}}

\def \dc {\text{dc}}
\def \ej {\text{ej}}
\def \exp {\text{exp}}

\def \iso {\text{iso}}
\def \v {\text{v}}
\def \sc {\text{sc}}
\def \ct {\text{ct}}
\def \hot {\text{hot}}
\def \cold {\text{cold}}
\def \z {\text{z}}
\def \eq {\text{eq}}
\def \tf#1 {\textcolor{blue}{ #1}}

\shorttitle{Shock Breakout from Stellar Envelopes: The relativistic limit}
\shortauthors{Tamar Faran \& Re'em Sari}

\begin{document}

\title{Shock Breakout from Stellar Envelopes: \\The relativistic limit}

\correspondingauthor{Tamar Faran}
\email{tamar.faran@mail.huji.ac.il}

\author{Tamar Faran}
\affiliation{Racah Institute of Physics, The Hebrew University of Jerusalem, Jerusalem 9190401, Israel}
\author{Re'em Sari}
\affiliation{Racah Institute of Physics, The Hebrew University of Jerusalem, Jerusalem 9190401, Israel}

\begin{abstract}
We calculate the observed luminosity and spectrum following the emergence of a relativistic shock wave from a stellar edges. Shock waves propagating at $0.6<\Gamma_\sh\beta_\sh$, where $\Gamma_\sh$ is the shock Lorentz factor and $\beta_\sh$ is its associated reduced velocity, heat the stellar envelope to temperatures exceeding $\sim 50$ keV, allowing for a vigorous production of electron and positron pairs. Pairs significantly increase the electron scattering optical depth and regulate the temperature through photon generation, producing distinct observational signatures in the escaping emission. Assuming Wien equilibrium, we find analytic expressions for the temperature and pair density profiles in the envelope immediately after shock passage, and compute the emission during the expansion phase. Our analysis shows that in pair loaded regions, photons are produced at a roughly uniform rest frame energy of $\sim 200$ keV, and reinforces previous estimates that the shock breakout signal will be detected as a short burst of energetic $\gamma$-ray photons, followed by a longer phase of X-ray emission. We test our model on a sample of low-luminosity gamma ray bursts using a closure relation between the $\gamma$-ray burst duration, the radiation temperature and the $\gamma$-ray isotropic equivalent energy, and find that some of the events are consistent with the relativistic shock breakout model. Finally, we apply our results to explosions in white dwarfs and neutron stars, and find that typical type Ia supernovae emit $\sim 10^{41}$ erg in the form of $\sim 1$ MeV photons.

\end{abstract}

\keywords{supernovae, gamma rays, shock breakout, shocks --- 
pair production}

\section{Introduction} \label{sec:intro}
Shock waves produced in stellar explosions such as supernovae (SNe) gain initial characteristic velocities of the order $v_\sh\sim \sqrt{2 E_\exp/M_\ej}$, where $E_\exp$ is the energy of the explosion and $M_\ej$ is the ejected mass. Typical combinations of $E_\exp$ and $M_\ej$ ($E_\exp\sim 10^{51}$ erg and $M_\ej\sim 1-10\,M_\odot$) produce shock waves moving at Newtonian velocities of $v_\sh \sim 2,000-7000$ km s$^{-1}$. As a shock wave propagates in the internal regions of the star, it decelerates while depositing its energy in the stellar material, but starts accelerating once it encounters the steeply decreasing density profile at the edge of the stellar envelope:
\begin{equation}\label{eq:density_profile}
    \rho = \rho_* \bigg(\frac{R_*-r}{R_*}\bigg)^n ~,
\end{equation}
where $R_*$ is the stellar radius, $r$ is the radial distance from the centre of the star, $\rho_*$ is a constant and the power law index $n$ is determined by the adiabatic nature of the envelope, where $n=3$ for radiative envelopes and $n = 3/2$ for convective envelopes. Depending on the explosion energy and the stellar radius, the shock may remain Newtonian or accelerate to relativistic velocities before the decreasing optical depth can no longer support it, and the shock breaks out of the star. At Newtonian velocities, $\beta_\sh\equiv v_\sh/c\ll 1$ where $c$ is the speed of light, the shock accelerates according to $v_\sh\propto \rho^{-0.19}$ \citep{sakurai60}, and may reach $5,000-10,000$ km s$^{-1}$ in Red Supergiant explosions, or $30,000-100,000$ km s$^{-1}$ in explosions of Wolf-Rayet stars \citep[e.g.,][]{Nakar2010,Faran2019b}. However, if the explosion is extremely energetic or takes place in a compact object like a white dwarf (WD) or a neutron star (NS), the shock may accelerate to mildly relativistic or even relativistic velocities, in which case the shock velocity evolves as $\Gamma_\sh\beta_\sh\propto t'^{-0.23n}$, where $\Gamma_\sh$ is the shock Lorentz factor and $\beta_\sh$ its corresponding reduced velocity and $t'$ is the lab frame time, which decreases to $0$ as the shock approaches the stellar surface. Hydrodynamic solutions for shocked relativistic flows can be found in e.g., \cite{Johnson1971}, \cite{BM76}, \cite{Sari2006} and \cite{Pan2006}.

As long as the shock propagates in the optically thick envelope of the star, the processes that decelerate the incoming flow (as seen in the shock frame) are governed by photon interactions such as Compton scattering, and under some conditions electron-positron pair-production, and the shock is therefore mediated by radiation. A deceleration zone of order a few photon diffusion lengths separates the upstream and the downstream, constituting the width of the shock \citep[see e.g.,][for a review]{Levinson2020}. A radiation mediated shock wave can survive in the stellar envelope as long as the optical depth of the medium ahead of it is much larger than its width. Once the diffusion time of photons in the downstream (propagating towards the upstream) becomes comparable to the dynamical time of the shock, the photons that decelerate the incoming flow can escape to the upstream and the shock can no longer be sustained. This implies that the shock breaks out of the star once the electron scattering optical depth ahead of it becomes smaller than $c/v_\sh$. Photons escaping the shock deceleration zone give rise to the \textit{shock breakout} emission, which is the first electromagnetic signal arriving from the explosion, typically in the UV to X-ray regime when the shock in Newtonian at breakout and in the $\gamma$-ray regime when it is relativistic. The breakout pulse is followed by the diffusion of photons from deeper regions inside the envelope that become optically thin due to expansion, often referred to as the `shock cooling' emission.

The observed properties of Newtonian shock breakout and shock cooling emission have been discussed extensively in the literature \citep[e.g., by][]{Matzner1999,Nakar2010,Rabinak2011,Faran2019b,Kozyreva2020,Goldberg2020} and are relatively well understood. However, the physics of shock breakout is more complicated in the relativistic regime. Once the temperature in the immediate downstream exceeds $\sim 50$ keV, pairs of electron and positron begin to form in significant amounts, increasing the number density of leptons and the electron-scattering optical depth. We denote the positron fraction with respect to protons as
\begin{equation}
    z_+ \equiv \frac{n_+}{n_\p}~,
\end{equation}
where $n_+$ and $n_\p$ are the positron and proton number densities, respectively. Following the above definition, the electron number density is $n_- = n_\p(1+z_+)$. A high multiplicity of pairs in the shocked envelope is expected to affect the observed signal, as will be shown in this work.

If advection of photons from the upstream dominates over photon generation in the downstream of the shock, like in subphotospheric internal shocks in GRBs, the shock is called `photon rich' \citep{Beloborodov2017,Ito2018,Lundman2018,Lundman2019}. In this paper we are concerned with the `photon poor' regime, in which most of the radiation is generated by local processes in the downstream. This scenario is appropriate for describing shocks propagating in undisrupted stellar envelopes. Photon poor radiation mediated shocks attain higher downstream temperatures and therefore pair creation becomes significant already at sub relativistic speeds, in contrast to $\Gamma_
\sh\beta_\sh>2$ in photon rich shocks \citep{Ito2018,Lundman2018}. 
Until now, only a handful of works have treated the problem of photon poor relativistic shock breakout under the effect of pair production. \cite{Budnik2010} and \cite{Ito2018} calculate the profiles of the Lorentz factor, pair density and temperature in the deceleration zone of relativistic radiation mediated shocks that propagate in a cold electron-proton plasma, using detailed numerical simulations. They also calculate the emerging photon spectrum in the immediate downstream, that would constitute the breakout emission. \cite{Granot18} develop an analytic model for a relativistic shock breakout from a stellar wind, and find predictions for the properties of the breakout signal, and \cite{Katz2010} find the radiation properties in the immediate downstream of a relativistic shock. All of the aforementioned studies treat only the breakout emission that originates from the shock deceleration region and the immediate downstream. However, the breakout signal is emitted over extremely short times scales (less than a millisecond) and is very faint, since only a small fraction of the shock energy is deposited in the layer from which the shock breaks out. Most of the energy is stored in deeper regions of the ejecta, and is emitted during the expansion phase of the envelope. The post breakout expansion dynamics are divided into a planar phase, before the radius of the envelope has doubled in size, and a spherical phase, in which the position of each fluid element evolves like $r\propto v\cdot t'$, where $v$ is the velocity. \cite{Nakar2012} calculate the observed luminosity and spectrum following a relativistic shock breakout during the planar and spherical phases, while restricting their discussion to cases in which the relativistic ejecta ends its post-breakout acceleration during the planar phase. They do not, however, solve for the exact temperature profile inside the envelope, and instead put an estimate on the post shock temperature.

In this work, we solve for the temperature and pair density profiles in the outer, post shock envelope, and calculate the properties of the escaping photons during planar and spherical expansion. We require that the medium is in a steady state of pair production and annihilation, assuming that photons form a Wien spectral distribution. The validity of this assumption is verified in a parallel publication (Faran \& Sari in prep, hereafter FS22), and is shortly discussed in Appendix \ref{app:Wien}. We are therefore able to perform a robust calculation of the radiation properties of the pair loaded plasma. We also consider cases in which the post breakout acceleration proceeds during the spherical phase, and find the terminal Lorentz factor acquired by fluid elements during spherical acceleration. This enables us to also treat the breakout of ultra-relativistic shock waves.

The structure of the paper is as follows: in section \ref{sec:equations} we summarize the equations governing our steady state model. In Section \ref{sec:initial_conditions} we compute the envelope properties immediately after shock passage, and describe the hydrodynamics of planar and spherical expansion in section \ref{sec:hydro}. The observed properties are calculated in Section \ref{sec:emission} and applied to different systems in Section \ref{sec:applications}. We summarize and discuss our results in Section \ref{sec:summary}.
\section{Pair-loaded equilibrium plasma}\label{sec:equations}
In this section we describe the equations governing the pair-loaded stellar envelope after the passage of a relativistic shock wave. We  use the following dimensionless parameters for the reduced temperature
\begin{equation}
    \theta \equiv \frac{k_\B T}{m_\e c^2}
\end{equation}
and the reduced photon energy
\begin{equation}
    x\equiv \frac{h \nu}{m_\e c^2}~,
\end{equation}
where $T$ is the temperature, $k_\B$ is the Boltzmann constant, $m_\e$ the electron mass, $h$ Planck's constant and $\nu$ the photon frequency, measured in the fluid rest frame. We use unprimed symbols to denote quantities measured in the rest frame of the fluid, while primed symbols denote quantities measured in the upstream (lab) frame.


\textbf{The total photon density:} In the optically thick regions of the stellar envelope, photon diffusion is unimportant relative to photon production. In addition, owing to the high temperatures expected in the downstream of relativistic shock waves, thermal equilibrium can only be achieved deep inside the envelope, where the temperatures are too low to allow for the existence of pairs. We therefore do not include photon escape nor absorption in our equations for the optically thick, pair loaded plasma (we do account for absorption in terms of the minimal energy below which the spectrum becomes a blackbody). In an optically thick plasma, the two main emission processes that dominate photon production are free-free and double-Compton (DC) emission, and the total photon density is given by:
    \begin{equation}\label{eq:ngamma_eq}
        n_\gamma = (\dot{n}^\ff+\dot{n}^\DC)t
    \end{equation}
where $\dot{n}^\ff$ and $\dot{n}^\DC$ are the photon production rates of free-free and double-Compton, respectively, and $t$ is the available time for photon production in the rest frame of the fluid. There are generally two main contributions to the free-free emission rate: electron-proton and electron-positron bremsstrahlung, where the latter is the dominant process in pair-loaded plasma. Since we are concerned with the limit of $1\ll z_+$, only the second term is important and the free-free photon emission rate is
\begin{equation}\label{eq:ndot_ff}
    \dot{n}^\ff = \frac{16}{3}\sqrt{\frac{2}{\pi}}\alpha c r_\e^2 n_+^2\theta^{-1/2}(2^{3/2}+4\theta)\Lambda g_\ff \,g_\WE^{1/2}\,,
\end{equation}
where $\alpha = 1/137$ is the fine structure constant, $\Lambda = \log{(\theta/x_m)}$, $g_\ff \simeq \log\Big(4\eta \sqrt{\theta/x_\m}\Big)$, $\eta \simeq \exp(-0.58)$ and $g_\WE \approx 1+3.76\theta+5.10\theta^2$ is a numerically fitted polynomial and is accurate to $0.06 \%$ \citep{Svensson84}. The photon energy $x_\m$ is the typical photon energy below which the spectrum becomes a Planckian. This frequency is determined by the requirement that the time-scale for Compton upscattering to the Wien peak is equal to the free-free absorption time scale. Since the plasma temperature is mildly relativistic, only one scattering is required to change the photon energy considerably. In the limit of $z_+\gg 1$, the lowest photon energy that can upscatter to the Wien peak is
\begin{equation}\label{eq:xm_ff}
    x^\ff_\m = \bigg(\frac{\alpha}{2\pi^{5/2}}\bigg)^{1/2} \bigg[\frac{\Lambda (1+\sqrt{2}\theta)}{g_\KN\, g_\WE^{1/2}}\bigg]^{1/2} \theta^{-3/4}(\lambda^3\, n_+)^{1/2}\,,
\end{equation}
where $g_\KN$ is the Klein-Nishina correction factor (see Section \ref{sec:initial_conditions}) and $\lambda = h/(m_\e c)$ is the Compton wavelength. For typical values of $\theta = 0.15$, $n_\p = 10^{15}$ cm$^{-3}$, $z_+ = 10^3$ and $g_\ff = 10$, we find that $\Lambda = \log(\theta/x^\ff_\m)\simeq12$.

The DC photon production rate is important only when the envelope is pair loaded, and produces photons at a rate of
\begin{equation}\label{eq:ndot_DC}
    \dot{n}^\DC = \frac{128}{3}\alpha c r_\e^2 2n_+ n_\gamma \theta^2 \Lambda g_\DC\,,
\end{equation}
where $g_\DC\simeq (1+13.91\theta)^{-1}$ and we assume a Wien spectral distribution. In pair dominated plasmas, the DC rate dominates over free-free at temperatures $\theta<0.1$ since it requires high photon densities with respect to the lepton density \citep[see][]{Svensson84}. When DC is the dominant photon production process, the expression for $x_\m$ at $z_+\gg 1$ is
\begin{equation}\label{eq:xm_DC}
    x^\DC_\m = \bigg(\frac{2}{\pi}\bigg)^{5/4}\alpha^{1/2}\Bigg(\frac{g_\DC}{g_\KN\, g_\WE^{1/2}}\Bigg)^{1/2} \theta^{5/4} \e^{1/2\theta} (\lambda^3\,n_+)^{1/2}\,.
\end{equation}
Typical values for regions dominated by DC of $\theta = 0.08$, $n_\p = 10^{15}$ cm$^{-3}$ and $z_+ = 10$ also give $\Lambda = \log(\theta/x^\DC_\m)\simeq12$.

\textbf{Pair production and annihilation:}
 Under the conditions that prevail in stellar envelopes, the main mechanism for pair production is photon-photon interaction \citep[e.g.,][]{Svensson82,Svensson84}. Two photons of energies $E_1$ and $E_2$ in their centre of momentum frame can produce an electron-positron pair if $\sqrt{E_1 E_2}\geq m_\e c^2$. Nevertheless, since the radiation in the downstream forms a spectral distribution that is close to Wien (see Appendix \ref{app:Wien}), there are enough `pair-producing' photons at $h\nu \sim m_\e c^2$ beyond the Wien peak even though $k_\B T< m_\e c^2$.
The shocked plasma is required to be in a state of pair equilibrium, namely, a steady state exists between pair production and annihilation. Pairs are produced by energetic photons created in local emission and scattering processes.
Given a Wien spectral distribution, photons create pairs at a rate of
    \begin{equation}\label{eq:ndot_pair_production}
      \dot{n}_\pm = \frac{\pi}{8} c \pi r_\e^2 n_{\gamma}^2  \theta^{-3} \e^{-2/\theta}~.
    \end{equation}
The reverse process of pair annihilation at non-relativistic temperatures ($\theta\ll 1$) occurs at a rate of
    \begin{equation}\label{eq:ndot_A}
        \dot{n}_\A  = n_- n_+ c \pi r_\e^2 ~.
    \end{equation}
This reaction releases two photons at an energy of $m_\e c^2$. However, these photons are not included in the photon balance equation since the net production rate of photons from pair annihilation and photon-photon ($\gamma \gamma$)-pair production is zero. Steady state between pair production and annihilation implies
\begin{equation}\label{eq:pair_balance_eq}
    \dot{n}_\pm =  \dot{n}_\A ~.
\end{equation}
\cite{Svensson82,Svensson84} find that there exists a critical temperature, which is dependent on the optical depth, above which pair annihilation is not able to balance pair production, and equilibrium cannot be achieved. However, the reason for this limit is a high temperature ($1<\theta$) correction to the pair annihilation cross section, which reduces the annihilation rate. At the temperature range relevant to this work a steady state solution always exists.

\textbf{Energy balance:} The downstream gas is heated by the shock wave and the energy is distributed among all plasma constituents and the photons. Pairs are thermally coupled to the protons and the radiation, and the following energy conservation equation holds
\begin{equation}\label{eq:energy_balance_eq}
    \epsilon \simeq \big(n_\gamma+n_-+n_+ + n_\p\big)3\theta~ m_\e c^2 ~,
\end{equation}
where $\epsilon$ is the internal energy density.

\textbf{Wien equilibrium:}
Wien equilibrium (WE) implies that the photon chemical potential satisfies $\mu_\gamma\ll 1$ and can be achieved through detailed balance for Compton scattering, and $\gamma\gamma$-pair production and annihilation.
For a Wien spectral density, a very simple relation between $n_\gamma$, $n_+$ and $\theta$ can be obtained from equations \eqref{eq:ndot_pair_production}, \eqref{eq:ndot_A} and \eqref{eq:pair_balance_eq}:
\begin{equation}\label{eq:WE}
    \frac{n_\gamma^2}{n_+ n_-} = \bigg[\frac{2 \theta^2}{K_2(1/\theta)}\bigg]^2 = \frac{8}{\pi}\theta^3\exp(2/\theta)
g_\WE^{-1}(\theta)~,
\end{equation}
where $K_2(x)$ is the modified Bessel function of the second kind. For non-relativistic temperatures ($\theta<1$), $n_\gamma/n_+ \gg 1$, while this ratio approaches unity when $\theta \gg 1$.
\section{The initial conditions after shock passage}\label{sec:initial_conditions}
In the following, we find the initial conditions in the envelope immediately after shock passage, as dictated by the shock jump conditions and by the model described in Section \ref{sec:equations}.
It is convenient to treat the stellar envelope as a sequence of successive shells, inside each the hydrodynamic and radiative properties, such as the density, velocity, and energy, as well as the photon density and temperature, are considered uniform. The depth inside the medium is characterised by the Lagrangian mass coordinate, $m$, defined to be the integrated mass from the edge of the medium ($m=0$) and increasing inwards up to the total ejecta mass $M_\ej$ at the base of the envelope.
In the pre-shocked envelope, each shell has a `pair-unloaded' scattering optical depth of
\begin{equation}
    \tau_\T = \sigma_\T n_\p d ~,
\end{equation}
where $\sigma_\T$ is the Thomson cross section and $d$ is the width of a shell, comparable to the distance from the base of the shell to the edge of the star. We denote the properties of the shell for which $\tau_\T=1$ with the subscript $_0$ and refer to it as the `pair-unloaded breakout shell', which would have been the shell from which the shock broke out in the absence of pairs. This definition implies that $\tau_\T = m/m_0$. For high enough shock velocities, pairs are produced in the downstream and the optical depth of each fluid element increases by a factor $1+2z_+$. We define the `pair-loaded' Compton optical depth as
\begin{equation}\label{eq:tau_tot}
    \tau_\tot = \tau_\T (1+2 z_+) g_\KN(\theta)
\end{equation}
where $g_\KN(\theta) \simeq (1+5\theta)^{-1}$ is the frequency averaged Klein-Nishina correction assuming Wien distribution and $\theta\leq1$ \citep{Svensson84}.

\begin{figure}
\centering
\includegraphics[width=1\columnwidth]{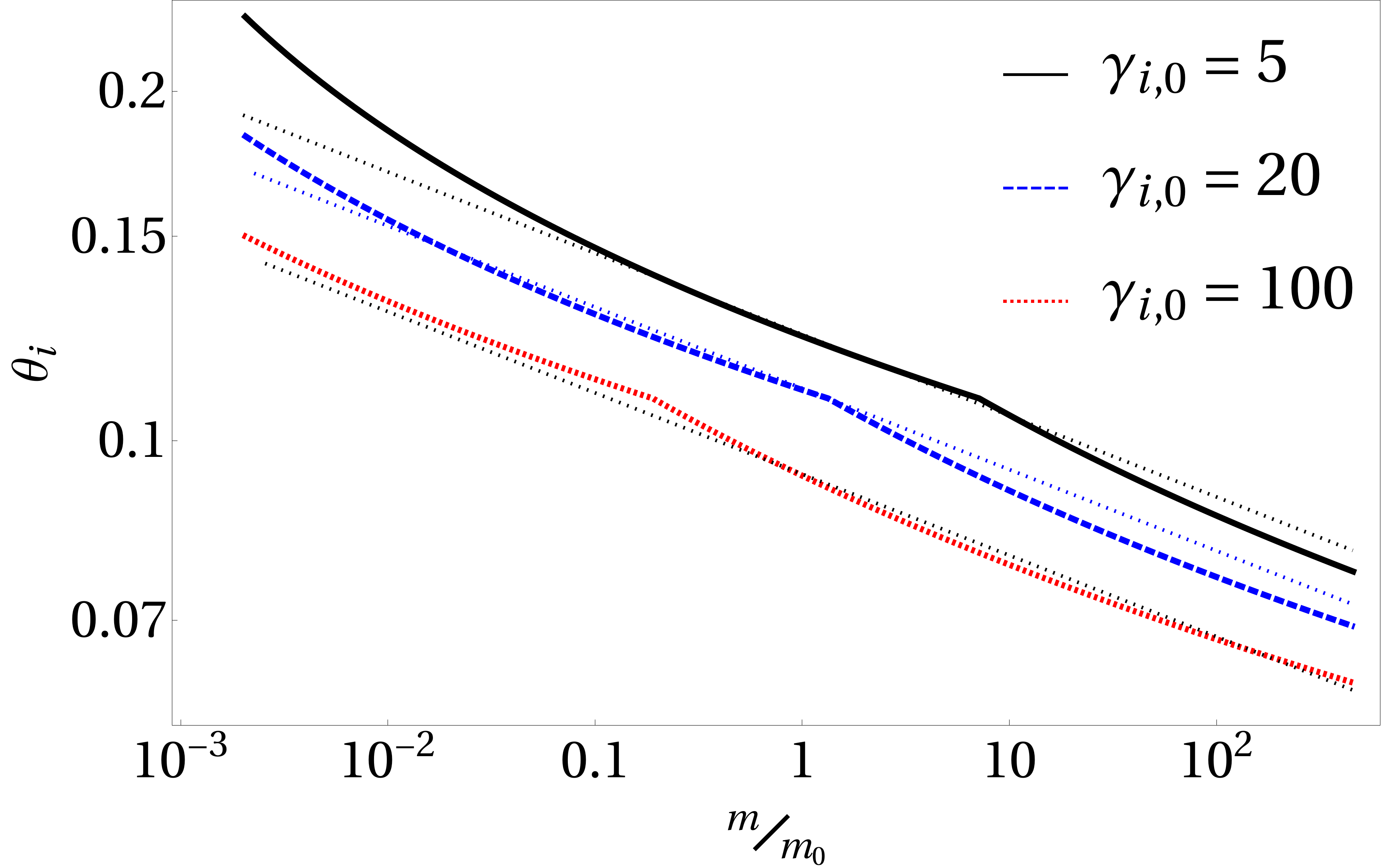}
\includegraphics[width=1\columnwidth]{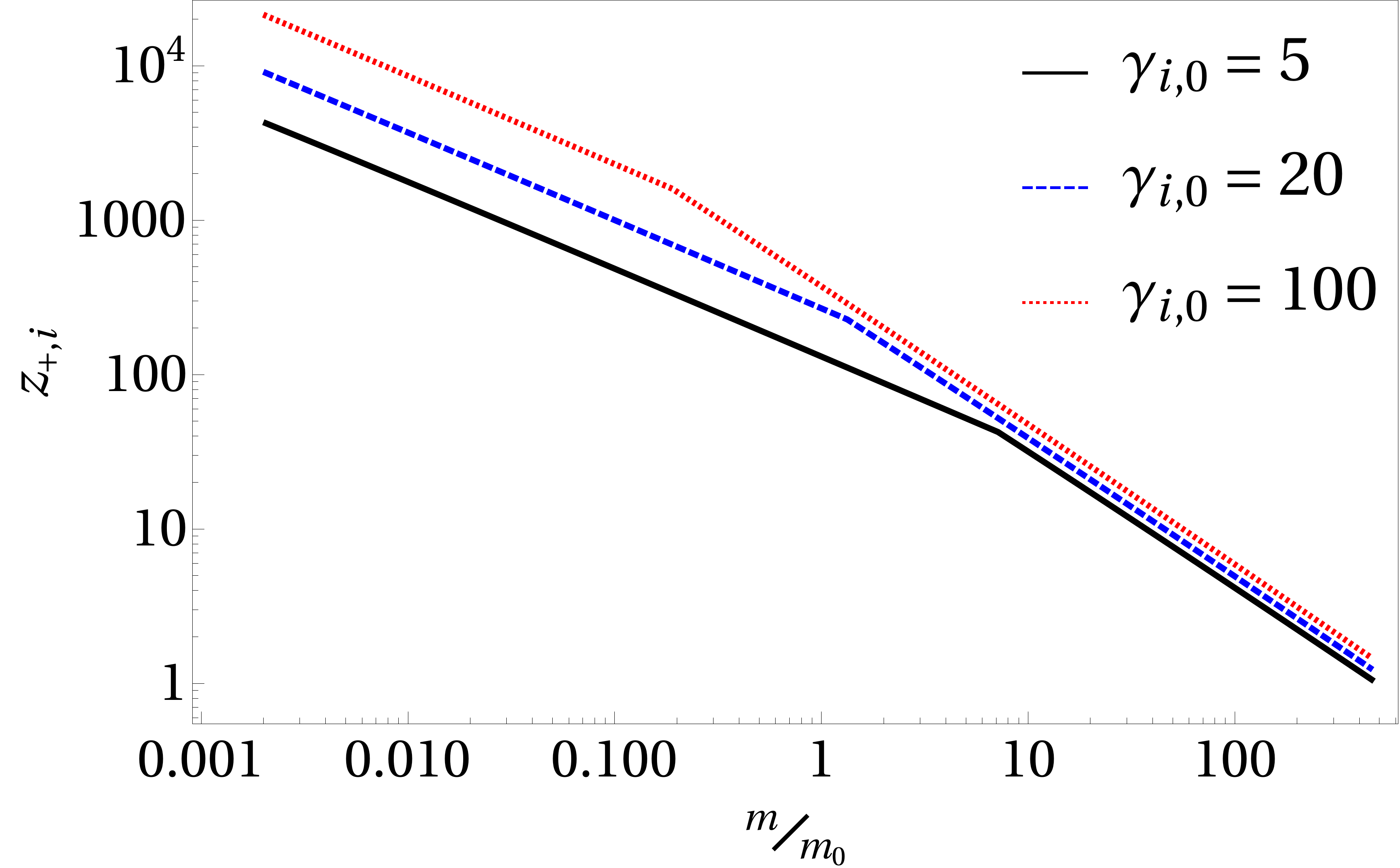}
\caption{Top: the temperature at $t_\i$ as a function of shell mass for $n=3$ and three different values of $\gamma_{\i,0}$, assuming WE and taking $\Lambda_\i \sim 10$ and $g_{\ff,\i}\sim 5$.  The dotted lines are fits to the numerical solution, and satisfy $\theta_\i\propto m^{-0.07}$. Bottom: the initial positron fraction for the same conditions as in the upper panel. The transition between free-free and DC domination occurs at $\theta_\i \sim 0.1$, and is visible by a break in the temperature and positron fraction curves. In the lower panel, the more notable change in the curve behaviour stems from the different dependence of $z_+$ on $\theta_\i$, as indicated in equations \eqref{eq:x_pos_ff}-\eqref{eq:x_pos_DC}. The trend of the temperature profiles with $\gamma_{\i,0}$ may seem counter-intuitive, with the higher Lorentz factors resulting in lower temperatures at a given mass coordinate. In FS22 we show that faster shock waves create more pairs and photons, which reduce the plasma temperature. Nevertheless, the breakout temperature itself increases with $\gamma_{\i,0}$, since the shock accelerates further and emerges from shallower shells.}\label{f:T_m_i}
\end{figure}

The conditions in the immediate downstream of the shock are calculated using the self-similar hydrodynamic solution obtained by \cite{Sari2006} for a planar converging shock wave. The pre- and post-shocked properties are denoted by the subscripts $_1$ and $_\i$, respectively. The proton number density of the undisturbed envelope as a function of $m$, following Eq \eqref{eq:density_profile} is
\begin{equation}
    n_{\p,1}(m) = n_0\bigg(\frac{m}{m_0}\bigg)^{\frac{n}{1+n}} ~,
\end{equation}
and the unshocked shell width is
\begin{equation}
    d_1(m) = \frac{1}{\sigma_\T n_0}\Big(\frac{m}{m_0}\Big)^{\frac{1}{n+1}} ~.
\end{equation}
The Lorentz factor, proton number density, energy density, shell width and dynamical time of a shell in the immediate downstream of the shock are
\begin{subequations}
\begin{alignat}{4}
   \gamma_\i(m) &= \gamma_{\i,0} \Big(\frac{m}{m_0}\Big)^{\frac{3-(2\sqrt{3})n}{2(1+n)}}\propto m^{-0.17}\,, \label{eq:gamma_i_m} \\
    n_{\p,\i}(m) &= 4 \gamma_\i n_{\p,1}\propto m^{0.58}\,,\\
    \epsilon_\i(m) &= 4\gamma_\i^2 m_\p c^2 n_{\p,1}\propto m^{0.40}\,,\\
  d_\i(m) &= \frac{d_1}{4 \gamma_\i}\propto m^{0.42}\,,\\
    t_\i(m) &= \frac{d_\i}{c} \,.
    \end{alignat}
\end{subequations}
respectively, where the proportionality power laws are calculated for $n=3$. This value of $n$ will be used in the remainder of this paper. The initial photon number density inside each shell is simply $n_{\gamma,\i} = \dot{n}_\gamma \cdot t_\i$.
We use equations \eqref{eq:ndot_ff} and \eqref{eq:WE} to obtain an expression for the initial pair fraction in WE, dominated by free-free photon production, assuming $1\ll z_+$ and $\theta \lesssim 1$:
\begin{equation}\label{eq:x_pos_ff}
    z_+^\ff(\theta_\i,x_{\m,\i}) =\bigg(\frac{m}{m_0}\bigg)^{-1} \frac{\pi}{2\alpha(\sqrt{2}+2\theta_\i)} \frac{e^{\frac{1}{\theta_\i}}\, \theta_\i^2}{\Lambda_\i g_{\ff,\i}}\,,
\end{equation}
where $g_{\ff,\i} \approx \Lambda_\i/2$, and the value of $\Lambda_\i$ is typically $\sim 10$. 
When $\theta_\i<0.1$ and $z_+ \gg 1$, DC is the dominant photon generation process, and using equations \eqref{eq:ndot_DC} and \eqref{eq:WE} we obtain
\begin{equation}\label{eq:x_pos_DC}
    z_+^\DC(\theta_\i,x_{\m,\i}) = \frac{\pi}{32 \alpha} \bigg(\frac{m}{m_0}\bigg)^{-1}\frac{1}{\theta_\i^2 \Lambda_\i g_{\DC,\i}}\,.
\end{equation}
Since the temperature is not expected to change significantly as long as the envelope is pair loaded, $z_+$ drops roughly linearly with the mass. Using equations \eqref{eq:energy_balance_eq}, \eqref{eq:WE} and \eqref{eq:x_pos_ff} or \eqref{eq:x_pos_DC}, we obtain implicit expressions for $\theta_\i(m)$ free-free and DC dmination:
\begin{equation}
    \frac{e^{2/\theta_\i} \theta_\i^{9/2}}{\Lambda_\i\, g_{\ff,\i}\,\sqrt{g_{\WE,\i}}(1+\sqrt{2}\theta_\i)} = \gamma_{\i,0}\frac{m_\p}{m_\e} \frac{\alpha}{3\sqrt{\pi}} \bigg(\frac{m}{m_0}\bigg)^{0.83} \,,
\end{equation}
and
\begin{equation}
    \frac{e^{1/\theta_\i} \theta_\i^{1/2}}{\Lambda_\i\, g_{\DC,\i}\sqrt{g_{\WE,\i}}} = \gamma_{\i,0}\frac{m_\p}{m_\e} \frac{2\sqrt{2}\alpha}{3^{1/3}\pi^{7/6}} \bigg(\frac{m}{m_0}\bigg)^{0.83} \,,
\end{equation}
respectively, where we accounted only for the photon contribution in Eq \eqref{eq:energy_balance_eq} since $n_\gamma\gg n_+,n_-,n_\p$. The exponent on the left hand side of both expressions implies a weak dependence of $\theta_\i$ on $\m$ and $\gamma_{\i,0}$ (and therefore on the shock Lorentz factor). While in shallower, hotter shells photon production is dominated by free-free emission, in more massive shells, $n_\gamma$ increases relative to $n_+$ and $n_-$, such that deep enough DC photon production rate dominates over free-free. In Figure \ref{f:T_m_i} we plot $\theta_\i(m)$ for three values of $\gamma_{\i,0}$, setting $\Lambda_\i= 10$. The transition between free-free and DC is visible as a break in the curves. The dependence of $\theta_\i$ on $m$ is evidently weak; over 6 orders of magnitude in mass, the temperature changes only by a factor of $\sim 3$, corresponding to a numerical fit of
\begin{equation}\label{eq:theta_i_m_fit}
    \theta_\i \propto m^{-0.07} \,,
\end{equation}
overplotted in Figure \ref{f:T_m_i} as dotted lines. It is evident that the fit to $\theta_\i$ describes the solution well for a wide range of $\gamma_{\i,0}$. When the medium is `pair loaded' ($z_+ \gg 1$), the pair plasma acts as a thermostat and prevents the downstream temperature from rising significantly above $\sim 100$ keV even when the shock is ultra-relativistic. This effect is attributed to the exponential sensitivity of the pair production rate to the temperature of the plasma (eq \ref{eq:ndot_pair_production}); even a slight increase in the temperature is followed by a copious production of pairs, which in turn generate photons at a rate proportional to $n_+n_-$ by free-free emission (or by double-Compton emission, in which case the rate is proportional to $n_-n_\gamma$). The energy is then distributed among all plasma constituents and the photons, and the temperature can increase only mildly. 
We use the numerical fit in Eq \eqref{eq:theta_i_m_fit} throughout this work instead of considering a constant temperature. In addition, due to the non-relativistic temperatures that characterize the plasma, we neglect relativistic KN effects.

The results of the steady state calculation are valid if the system is able to reach equilibrium within a dynamical time. We find that $t_\eq/t_\dyn \sim 1/\tau_\tot$, where $t_\eq = n_+/\dot{n}_\A= n_+/\dot{n}_+$ is the time scale to reach equilibrium. In opaque shells $\tau_\tot>1$, and equilibrium is typically achieved.

\section{Relativistic Expansion Dynamics}\label{sec:hydro}
\subsection{Planar Expansion, $\lowercase{t}_\i<\lowercase{t}<\lowercase{t_\s}$}\label{sec:planar_hydro}
We apply the self-similar solution obtained by \cite{Pan2006} for the hydrodynamics of a freely-expanding ejecta after shock breakout, for the initial expansion phase. Planar expansion governs the dynamics until the spherical time, $t'_\s\simeq R_*/c$, where $t'$ is the time measured in the upstream frame. The solution assumes that $ \epsilon\gg\rho c^2$, and is therefore valid as long as the fluid is hot. The evolution of the proton density, the internal energy and the Lorentz factor of a fluid element, respectively, obey:
\begin{subequations}
\begin{alignat}{3}
n_\p(t_\i<t) &= n_{\p,\i}\bigg(\frac{t}{t_\i}\bigg)^{-1}           \label{eq:n_aplanar_hot}\\
\epsilon(t_\i<t) &= \epsilon_\i\bigg(\frac{t}{t_\i}\bigg)^{-4/3}\label{eq:epsilon_aplanar_hot}\\
\gamma(t_\i<t) &= \gamma_\i\bigg(\frac{t}{t_\i}\bigg)^{1/\sqrt{3}}~,\label{eq:gamma_aplanar_hot}
\end{alignat}
\end{subequations}
where $t$ is the time measured in the rest frame of the fluid. After a fluid element has been shocked, it expands and accelerates as its internal energy is converted to bulk kinetic energy. The acceleration ceases once $\epsilon\sim \rho c^2$, which implies a final Lorentz factor of\begin{equation}\label{eq:gamma_f_planar}
    \gamma_\f = 1.96 \gamma_\i^{\sqrt{3}+1}~,
\end{equation}
where the prefactor is calculated for $n=3$ \citep{Pan2009}. \cite{Pan2009} also found the hydrodynamic solution of the flow during the cold planar phase. In terms of the rest frame time, the evolution of the energy and number densities are the same as in Eq \eqref{eq:n_aplanar_hot}--\eqref{eq:epsilon_aplanar_hot}, while the Lorentz factor remains constant.
The total photon number in a relativistically expanding plasma is approximately constant in time, i.e., the number of new  photons generated by free-free or DC emission is a decreasing function of time (this is true during both planar and spherical phases). All photons are therefore effectively produced at $t_\i$, implying that the evolution of the temperature and photon energy is adiabatic and the shape of the spectrum is conserved. Therefore, if a shell had a Wien spectral distribution at $t_\i$, it is maintained during the entire planar evolution. The total photon density and the temperature evolve in time according to
\begin{subequations}
\begin{alignat}{3}
    n_\gamma(t_\i<t)  &= n_{\gamma,\i}\bigg(\frac{t}{t_\i}\bigg)^{-1} \,\label{eq:n_ph_planar}\\
    \theta(t_\i<t) &= \theta_\i \bigg(\frac{t}{t_\i}\bigg)^{-1/3}~,
\label{eq:theta_t_planar}
\end{alignat}
\end{subequations}
where the time evolution of $x_\m$ is similar to that of $\theta$, meaning that $\Lambda \equiv \log(\theta/x_\m)$ keeps its initial value.
Steady state between pair production and annihilation is maintained also during the expansion phase. Using equations \eqref{eq:WE} and \eqref{eq:n_ph_planar}--\eqref{eq:theta_t_planar}, we find that the pair fraction drops exponentially with $\theta$ and therefore with time:
\begin{equation}\label{eq:n_pos_t}
    n_+(t_\i<t) = \sqrt{\frac{\pi}{8}} n_{\gamma,\i} \theta_\i^{-\frac{3}{2}} \bigg(\frac{t}{t_\i}\bigg)^{-1/2} e^{-\frac{1}{\theta_\i}\big(\frac{t}{t_\i}\big)^{1/3}} ~.
\end{equation}
The exponential dependence of the positron density on time implies that pairs quickly annihilate, significantly reducing the optical depth of pair-dominated shells within just a few dynamical times.
\subsection{Spherical expansion, $\lowercase{t_\s}<\lowercase{t}$}
At $t'>t'_\s$, where $t'_\s = R_*/c$, the expansion becomes spherical and the radius of the ejecta increases like $R\simeq c t'$. In Appendix \ref{app:spherical_dynamics}, we derive the spherical hydrodynamic solution for a relativistically expanding flow using the generalized Riemann invariant \citep{OrenSari2009}. We divide our discussion below to hot and cold spherical dynamics.
\subsubsection{Hot Spherical Dynamics}
We start with the evolution of the flow where $\epsilon\gg\rho c^2$, which applies to shells that have not had time to cool before $t_\s'$. The fluid's Lorentz factor, energy density and number density satisfy
\begin{subequations}
\begin{alignat}{3}
    &\frac{d \gamma}{dt'} = \frac{\gamma}{t'} ~,\label{eq:gamma_t_sph_hot}\\
    &\frac{d \epsilon}{dt'} = -4 \frac{\epsilon}{t'} ~,\label{eq:epsilon_t_sph_hot}\\
    &\frac{d n}{dt'} = -3\frac{n}{t'}\,,\label{eq:n_t_sph_hot}
    \end{alignat}
\end{subequations}
written in terms of the lab frame time due to a non-trivial conversion between rest frame and lab frame time\footnote{During spherical acceleration, the conversion between lab frame and rest frame time is $t = \frac{t'_\s}{\gamma_\s}\Big(\ln{\frac{t'}{t'_\s}}+1\Big)$.}. The temporal evolution of $\gamma$ (Eq \ref{eq:gamma_t_sph_hot}) suggests that as long as a fluid element is hot, its lab frame width, $\Delta'$, remains approximately constant:
\begin{equation}
    \Delta' = \Delta'(t'_\s) + \frac{c t'}{\gamma^2} \simeq \Delta'(t'_\s) ~,
\end{equation}
where $\Delta'(t'_\s) = \frac{c t'_\s}{\gamma_\s^2}$ is the width of the shell at $t'_\s$ and $\gamma_\s$ is its Lorentz factor at that time.
A fluid element does most of its expansion until $t'_\s$, where during the hot spherical phase it can only roughly double its width. Since shells maintain a constant width, their dynamical time corresponds to the time it takes the density to change: $t'_\dyn \simeq r/c$.
The fact that the width of hot shells remains constant, limits the ability of photons to cross them. The distance between the shell's external boundary and a photon emitted at the internal boundary is $\sim \frac{\Delta'_\s}{2}\big(1+\frac{t'_\s}{t'}\big)$,
which approaches $\Delta_\s'/2$ in the limit $t'_\s \ll t'$.
Photons manage to cross roughly halfway through the width of the shell during the spherical acceleration phase.\footnote{The case of $\gamma\propto t'$ corresponds to a constant  proper acceleration, where the lab frame coordinates are represented by Rindler coordinates. In this case, it is known that there exists a minimal distance between a photon and an accelerating observer (the edge of a shell, in our case), beyond which a photon will never be able to reach the observer.} This will not have any effect on the escaping radiation from $\tau_\tot=1$. However, the acceleration of the flow continues only as long as radiation is coupled to the electrons and the force exerted on an electron by the photons is sufficient to accelerate it to the bulk Lorentz factor of the radiation field (the Lorentz factor of the frame in which the radiation is isotropic). Therefore, work done on a single electron during the dynamical time $r/c$ needs to satisfy $\gamma m_\e c^2 \leq \sigma_\T \frac{c}{2 \gamma^2}  \frac{r}{c} \epsilon'$. This inequality imposes a lower limit on the Lagrangian mass that can be accelerated to the required Lorentz factor. Nevertheless, in Appendix \ref{sec:decoupling} we find that shells reach their terminal Lorentz factor before electrons and photons decouple, in which case neither the dynamics nor the emitted radiation are affected.
\subsubsection{Cold Spherical Dynamics}
Once a shell has exhausted most of its internal energy, it stops accelerating. The transition to the cold spherical phase implies that the width of fluid elements starts increasing with time:
\begin{equation}
    \Delta'= \Delta'(t'_\s) + \frac{ct'}{\gamma_\f^2} \simeq \frac{ct'}{2\gamma_\f^2}\,,
\end{equation}
where $\gamma_\f$ is the terminal Lorentz factor acquired by a fluid element through spherical acceleration. As shown in Appendix \ref{app:spherical_dynamics}, the evolution of $\epsilon$ and $n$ in terms of lab frame time remains the same as in the hot spherical phase, given by Eq \eqref{eq:gamma_t_sph_hot}--\eqref{eq:n_t_sph_hot}.
Using Riemann invariants, we show in Appendix \ref{app:spherical_dynamics} that $\gamma_\f$ is equal to \footnote{We note the mistake in the derivation of \cite{Yalinewich2017}, who assume that the width of a shell increases during the spherical phase, and hence derive an incorrect expression for $\gamma_\f$.}
\begin{equation}\label{eq:gamma_f_sph}
    \gamma_\f = \gamma_{\i,*}^{1+\sqrt{3}}\bigg(\frac{\gamma_\i}{\gamma_{\i,*}}\bigg)^{2+\frac{2}{3n}}  \,,
\end{equation}
where $\gamma_{\i,*}$ is the initial Lorentz factor of the shell that reaches its final Lorentz factor at $t_\s$, and is given by Eq \eqref{eq:gs_gf}.
Similarly to the relativistic planar phase, thermal coupling between electrons and photons cannot increase since the number of photons remains constant, and the evolution of the photon density and temperature obeys
\begin{subequations}
\begin{alignat}{3}
    n_\gamma(t'_\s<t')  &= n_{\gamma,\i}\bigg(\frac{t_\s}{t_\i}\bigg)^{-1}\bigg(\frac{t'}{t'_\s}\bigg)^{-3} ~,\\
    \theta(t'_\s<t') &= \theta_\i\bigg(\frac{t_\s}{t_\i}\bigg)^{-1/3} \bigg(\frac{t'}{t'_\s}\bigg)^{-1}\,.\label{eq:theta_spherical}
\end{alignat}
\end{subequations}
Equations \eqref{eq:gamma_t_sph_hot} and \eqref{eq:theta_spherical} imply that during the hot spherical phase, the observed (lab frame) temperature of a fluid element remains constant in time, while decreasing as $t'^{-1}$ in the cold spherical phase.


\section{Observed temperature and luminosity}\label{sec:emission}
\subsection{Planar Phase Emission}\label{sec:planar_emission}
The observed emission originates in the shell from which photons can effectively escape over a dynamical time, which in relativistic flows means that its optical depth satisfies $\tau_\tot = 1$ (although, we shall see that in some cases this statement needs to be refined). We call this shell the luminosity shell and denote its properties with the subscript $_\tr$. The shell from which the shock breaks out of the envelope is denoted as the `breakout shell', and is the source of the first photons that are emitted from the explosion. The breakout shell satisfies $\tau_\tot = 1/\beta_\d \simeq 3$, where $\beta_\d$ is the velocity in the immediate downstream in the shock frame. From Eq \eqref{eq:tau_tot}, we see that a shell of mass $m$ becomes transparent to photons when its pair fraction reduces to
\begin{equation}\label{eq:x_tr}
    z_{+,\tr} =\frac{1}{2}\bigg[\bigg(\frac{m}{m_0}\bigg)^{-1} -1 \bigg]\,,
\end{equation}
neglectiong Klein-Nishina effects.
Since $n_+$ drops exponentially with time (Eq \ref{eq:n_pos_t}), the pair loaded optical depth decreases rapidly, while the pair-\textit{unloaded} optical depth remains constant in time. Therefore, the optical depth can change only due to pair annihilation (and to a change in $\theta$ through KN effects, although this contribution is negligible since $\theta \ll 1$), such that when $z_+$ drops below $1$, the optical depth remains constant in time until the onset of the spherical phase.
As a result, the deepest shell that contributes to the planar phase emission is the unloaded breakout shell \citep[however, see][]{Faran2019b}.

In order to find the temperature of each shell as it becomes optically thin, $\theta_\tr$, we use Equations \eqref{eq:WE}, \eqref{eq:x_pos_ff}, \eqref{eq:n_ph_planar}, and \eqref{eq:x_tr} in the limit $m \ll m_0$, $1\ll z_+$ and $\theta_\i \ll 1$: 
\begin{equation}\label{eq:theta_tr}
    \theta_\tr = \ln \Bigg[\frac{e^{2/\theta_\i}\theta_\i^{7/2}}{\Lambda_\i\, g_{\ff,\i}\theta_\tr^{3/2}}\frac{\pi}{\sqrt{2}\alpha}\Bigg]^{-1}~.
\end{equation}
The explicit dependence of $\theta_\tr$ on $m$ cancels out, and only exits through $x_{\m,\i}$ and $\theta_\i$. Since the ratio $\theta_\i/x_{\m,\i}$ and $\theta_\i$ are weak functions of $m$, the rest frame temperature of the photons is almost uniform. We find that a shell becomes transparent at $t_\tr(m) \sim 5\,t_\i(m)$, i.e., after $\sim 2$ multiplications of the dynamical time, so that $\theta_\tr/\theta_\i \sim 5^{-1/3}\simeq 0.6$. We adopt this result to account for the time shells become transparent. Using equations \eqref{eq:theta_i_m_fit} and \eqref{eq:theta_tr}, one finds that $\theta_\tr$ ranges between 0.12 and 0.08, over 2 orders of magnitude in mass. Nevertheless, while the escaping photon energy in the rest frame of the fluid is roughly constant, the observed energy is not, due to Lorentz boost (for shells that cooled during the planar phase, $\gamma_\f \propto m^{-0.48}$).

Shells cool rapidly from the bottom towards the edge of the envelope, where the `cooling shell' satisfies $\epsilon = \rho c^2$ and evolves as $m_\c(t) \propto t^{-10.2}$. Naively, the outermost shell that has reached its final Lorentz factor by the end of the planar phase is located at $m_{\c,\pl} =  0.09\,m_0\,(n_0 \sigma_\T R_*)^{-1.74} \gamma_{\i,0}^{8.25}$.
Since $n_0\sigma_\T R_* \gg1$, a high Lorentz factor is required to keep the unloaded breakout shell hot at the end of the planar phase. Moreover, acceleration is also limited by the time a shell becomes transparent; once photons escape, radiation can no longer provide the pressure required for acceleration. Therefore, if a shell becomes transparent before reaching $\gamma_\f$ predicted by Eq \eqref{eq:gamma_f_planar}, its Lorentz factor only increase by a factor of $\sim 2.5$ before it becomes transparent. Although shells release more of their internal energy in this case, the energy in the observer frame is lower than it could have been had acceleration continued, due to a smaller Lorentz boost.

The condition that the unloaded breakout shell reaches its terminal Lorentz factor before becoming transparent sets an upper limit on the value of $\gamma_{\i,0}$, of $\gamma_{\i,0} \lesssim 1.96^{-1/\sqrt{3}} 5^{1/3} \simeq 1.2$. This result is independent of the physical parameters of the system.

In the observer frame, the arrival time of photons is measured in terms of $t_\obs$ -- the time difference between the detection of the first photon that escaped the ejecta, and a photon that reached the observer at time $t'$. The relation between the observed and lab frame times is $t_\obs = t'/(2\gamma^2)$, where $\gamma$ corresponds to the Lorentz factor of the shell from which the observed photon originated. Since the unloaded breakout shell is the deepest shell that can be exposed during the planar phase, its properties define the observed spherical time:
\begin{equation}\label{eq:t_s_obs}
    t^{\obs}_\s=  \frac{R_*/c}{2\gamma_{0,\tr}^2}\,,
\end{equation}
where $\gamma_{0,\tr}$ is the Lorentz factor of the unloaded breakout shell when it becomes optically thin, which is equal to $\gamma_\f$ defined in Eq \eqref{eq:gamma_f_planar} if it is cold, or to $\sim 2.5\gamma_{\i,0}$ if it becomes transparent before reaching $\gamma_\f$. The coordinate of the luminosity shell is determined by the condition $t_\obs \simeq 5\,t_\i/\gamma_\tr$, which gives
\begin{equation}\label{eq:mtr_tobs}
    m_\tr \propto \begin{cases}
        t_\obs^{1.67} &, t^\obs<t^\obs_\c \\
        t_\obs^{1.11} &, t^\obs_\c<t^\obs<t^\obs_0\,,
    \end{cases}
\end{equation}
where $t^\obs_0$ is the observed time at which the unloaded breakout shell becomes transparent and $t_\c^\obs$ is the observed time at which $m_\tr = m_\c$:
\begin{equation}\label{eq:tc_obs}
    t_\c^\obs = 0.006\frac{R_*}{c}(n_0\sigma_\T R_*)^{-1} \gamma_{\i,0}^{1.44}\,.
\end{equation}
The coordinate of the shell satisfying the above condition is equal to $m_{\c,\tr}\simeq 0.002\,m_0 \gamma_{\i,0}^{5.75}$. The result that all pair loaded shells become transparent at $\sim 5\, \t_\i$, implies that the final Lorentz factor of this shell is $\gamma_{\f,\c} \simeq 1.96^{-\frac{1}{\sqrt{3}}}\times 5^{\frac{1+\sqrt{3}}{3}}\simeq 3$.

During the planar phase, the observed signal from different shells is mixed due to light travel time and beaming effects. The luminosity and typical photon energy of a single shell of mass $m$ is
\begin{equation}\label{eq:Lobs_m}
\begin{split}
    L^{\obs}(m)& = \frac{E_\tr}{(2R_*/c)/\gamma_\tr}\cdot\frac{1}{\big[\gamma_\tr (1-\beta_\tr\,\cos \phi)\big]^2}\\
    &\simeq \frac{\gamma_\tr E_\tr}{(2R_*/c)/\gamma_\tr^2}\frac{1}{\Big[t_\phi^\obs\big(\frac{2 \gamma_\tr^2}{R_*/2c}\big)+1\Big]^2}
    \end{split}
\end{equation}
and
\begin{equation}\label{eq:nu_m}
    \nu'(m) \simeq \frac{\gamma_\tr\nu}{t_\phi^\obs\big(\frac{2 \gamma_\tr^2}{R_*/2c}\big)+1}\,,
\end{equation}
respectively, where $t^\obs_\phi = \frac{R_*}{c}(1-\cos{\phi})$ is the difference in arrival times of photons emitted at angles $\phi$ and $0$ relative to the line of sight, and we applied the ultra-relativistic approximation of $\beta\simeq 1-\frac{1}{2\gamma^2}$. $L^\obs(m)$ and $\nu'(m)$ are therefore effectively constant on a timescale of $R_*/(2c\gamma_\tr^2)$, where most of the energy originates from inside a beaming cone of opening angle $1/\gamma_\tr$. At timescales $t_\obs>\frac{R_*/c}{\gamma_\tr^2}$, the observed luminosity and spectra of each shell are governed by high latitude emission from angles $\phi>1/\gamma_\tr$, and their time evolution satisfies $L^\obs(m)\propto t_\obs^{-2}$ and $\theta^\obs(m)\propto t_\obs^{-1}$, as seen from equations \eqref{eq:Lobs_m}-\eqref{eq:nu_m} \citep[see also][]{Kumar2000}.
The above derivation assumes that the Lorentz factor of the shell does not increase once photons start leaking out, and that its radius remains constant. The observed luminosity at $t_\obs$ is the sum over the contribution from all transparent shells, and is governed by the most massive transparent shell.

\begin{figure*}
    \centering
    \includegraphics[width = 0.75 \textwidth]{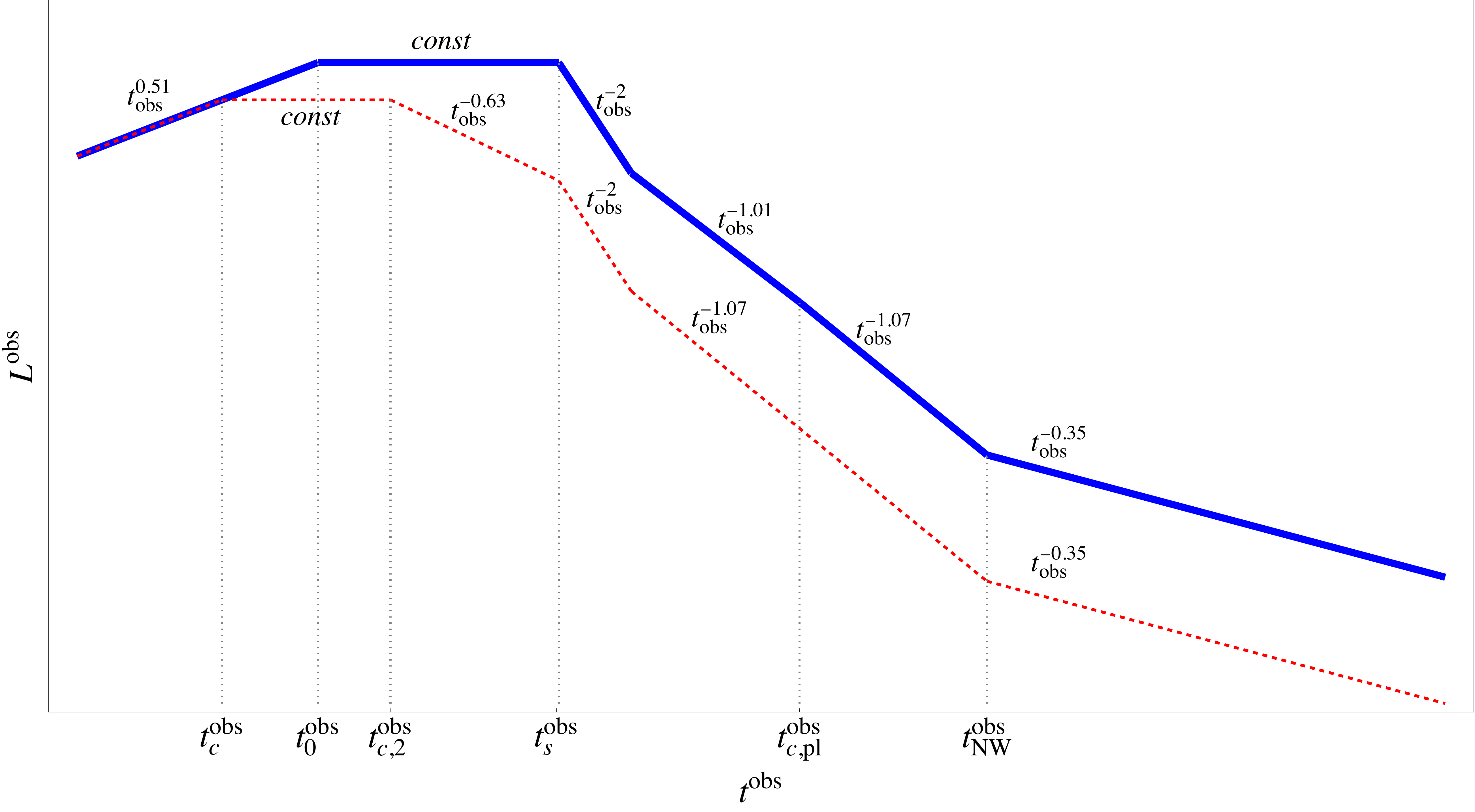}
    \caption{Schematic evolution of the observed luminosity. Blue: the case in which the unloaded breakout shell is still hot at the end of the planar phase. Dashed red: the case in which the unloaded breakout shell cools before $t_\s^\obs$. For real progenitor parameters, the values of typical time scales for the two cases are different, and their convergence in this figure is merely for illustrative purposes.}
    \label{fig:Lobs_schematic}
\end{figure*}

\begin{figure*}
    \centering
    \includegraphics[width = 0.75 \textwidth]{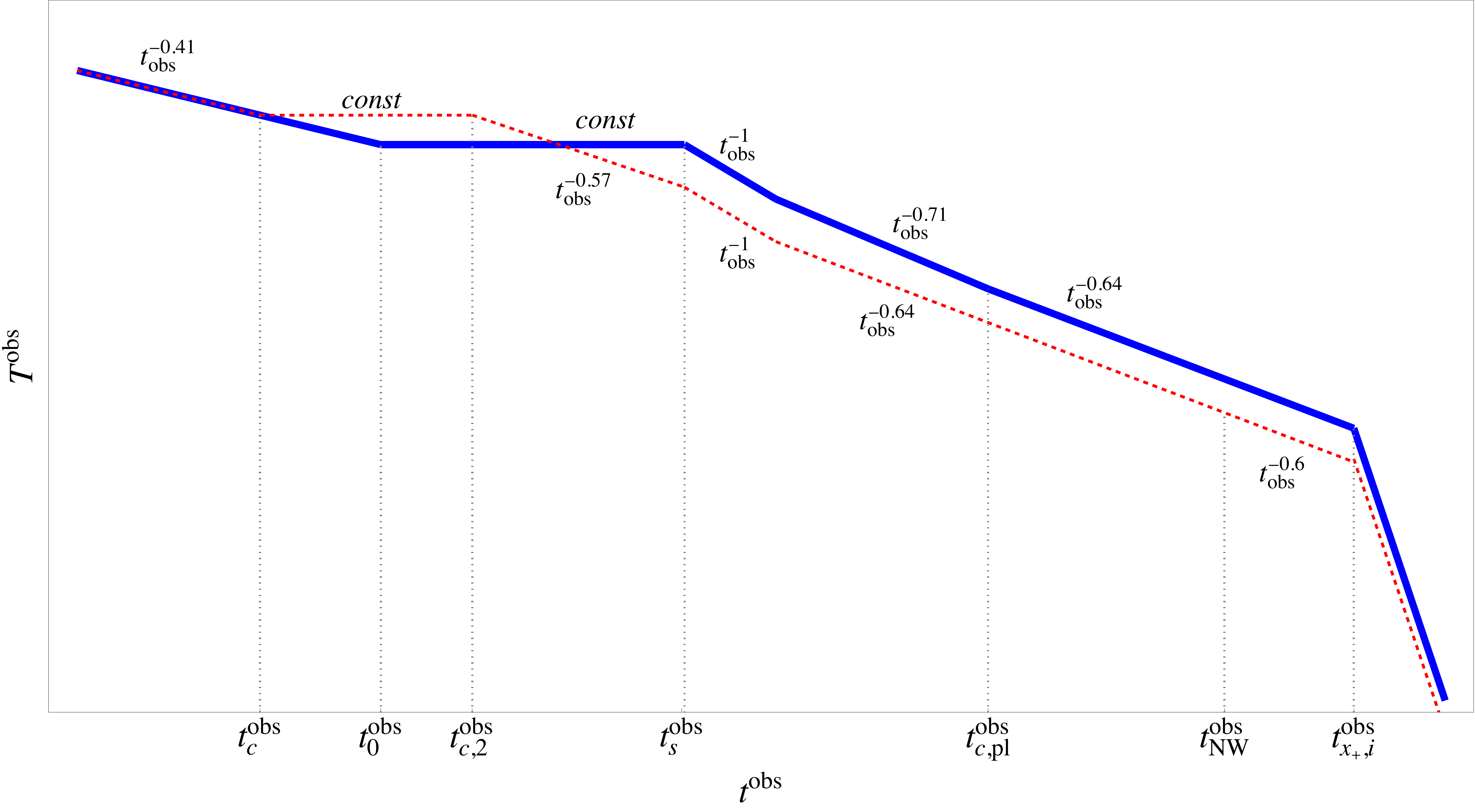}
    \caption{Same as Figure \ref{fig:Lobs_schematic} for the observed temperature.}
    \label{fig:Tobs_schematic}
\end{figure*}

Equations \eqref{eq:mtr_tobs} and \eqref{eq:Lobs_m} give the time evolution of the planar phase luminosity. If the unloaded breakout shell ends its acceleration before it becomes transparent, we refer to the planar phase as `cold'. The luminosity in this case evolves as
\begin{equation}\label{eq:Lobs_planar_cold}
\begin{split}
    L^{\obs}|_\text{cold}& = \frac{\gamma_\tr E_\tr}{R_*/(2\gamma_\tr^2 c)}\simeq  \, 0.6\,\frac{\gamma_{\i,\c}^{3(1+\sqrt{3})} E_{\i,\c}}{R_*/c}\times \\
    \times &
    \begin{cases}
    \Big(\frac{t_\obs}{t^\obs_\c}\Big)^{0.51}, & t_\obs< t^\obs_{\c}\\
    \text{const} &, t^\obs_{\c}<t_\obs< t^\obs_{\c,2}\\
    \Big(\frac{t^\obs}{t^\obs_{\c,2}}\Big)^{-0.63} & , t^\obs_{\c,2}< t_\obs<t^\obs_\s\\
    \end{cases} 
\end{split}
\end{equation}
where $t_{\c,2}^\obs = R_*/(2 c\gamma_{\c,\f}^2)$, $\gamma_{\c,\f}$ is the terminal Lorentz factor of the shell located at $m_{\c,\tr}$, $\gamma_{\i,\c}$ is its initial Lorentz factor and $E_{\i,\c}$ is its initial energy. At $t_\obs>t_\c^\obs$ the luminosity originating from $\tau_\tot=1$ starts decreasing, so that the main contribution to the luminosity is still coming from $m_{\c,tr}$, and the luminosity is therefore constant at $t_\c^\obs<t_\obs<t_{\c,2}^\obs$. Afterwards, the luminosity decreases as it is governed by more internal shells that enter their spherical phase at $t_\obs$. We note that at $t_\c^\obs<t_\obs<t_\s^\obs$ the observed emission is not governed by emission from $\tau_\tot = 1$. Following the above discussion, the evolution of the observed temperature is
\begin{equation}\label{eq:Tobs_planar_cold}\begin{split}
    \theta^\obs|_\text{cold} &= \gamma_\f\, \theta_\tr \simeq 1.2\,\gamma_{\i,\c}^{1+\sqrt{3}}\,\theta_{\i,\c} \\
    &\times\begin{cases}
    \Big(\frac{t_\obs}{t^\obs_\c}\Big)^{-0.41}\,, & t_\obs<t^\obs_\c\\
    \text{const}\,, &
    t^\obs_\c<t_\obs<t^\obs_{\c,2} \\
    \Big(\frac{t_\obs}{t^\obs_{\c,2}}\Big)^{-0.57}\,, & t^\obs_{\c,2}<t_\obs<t^\obs_\s\,.
    \end{cases}
\end{split}
\end{equation}
If the unloaded breakout shell has not exhausted its internal energy before it is exposed, the planar phase is referred to as `hot'. In this case, the unloaded breakout shell governs the luminosity and temperature during $t^\obs_0<t_\obs<t^\obs_\s$, where by $t^\obs_0$ is the time it becomes transparent, which again results in constant luminosity and temperature phases. The evolution of the luminosity and temperature in the hot planar phase is:
\begin{equation}\label{eq:Lobs_planar_hot}
\begin{split}
    L^{\obs}|_\text{hot}&= \frac{\gamma_\tr E_\tr}{R_*/(2\gamma_\tr^2 c)}\simeq  \, 30\,\frac{\gamma_{\i,0}^4 m_\p c^2}{R_*/c} \frac{R_*^2}{r_\e^2}\times\\
    &\times
    \begin{cases}
    \Big(\frac{t_\obs}{t^\obs_0}\Big)^{0.51} &, t_\obs< t^\obs_0\\
    \text{const} & , t_0^\obs< t^\obs<t^\obs_\s \,.
    \end{cases} 
    \end{split}
\end{equation}
and
\begin{equation}\label{eq:Tobs_planar_hot}
\begin{split}
    \theta^\obs|_\text{hot} &= \gamma_\tr\, \theta_\tr \simeq 1.5\,\gamma_{\i,0}\theta_{\i,0}\\
    &\times\begin{cases}
    \Big(\frac{t_\obs}{t^\obs_0}\Big)^{-0.41}\,, & t_\obs<t^\obs_0\\
     \text{const}\,, & t^\obs_0<t_\obs<t^\obs_\s\,,
    \end{cases}
    \end{split}
\end{equation}
where $E_{\i,0} \simeq \gamma_{\i,0}m_\p c^2 R_*^2/r_\e^2$ is the initial energy of the unloaded breakout shell.
A schematic evolution of the luminosity and temperature in the hot and cold scenarios is shown in Figures \ref{fig:Lobs_schematic}-\ref{fig:Tobs_schematic}.

Each shell contributes to the observed spectrum of $\nu F_\nu^\obs$ a Wien shaped spectral density with a peak photon energy of $4 \gamma_\tr \theta_\tr$. The observed spectrum is calculated by taking the characteristic peak photon energy of each transparent shell weighed by its relative contribution to the luminosity of the pulse. In the cold planar phase (where the unloaded breakout shell cools before $t_0^\obs$), the spectrum at $t_{\c,2}<t_\obs^\obs<t_\s^\obs$ peaks at the temperature of the shell entering its spherical phase at $t_\obs$, and has an exponential cutoff at higher frequencies. Below the peak, contribution to the spectrum comes from all exposed shells, and at $\nu'<\nu_0'$ mainly from the the unloaded breakout shell, where $\nu_0'$ is its peak photon energy: 
\begin{equation}\label{eq:planar_spectrum_cold}
    \nu F_\nu^\obs|_\text{cold} \propto \begin{cases}  
    \nu'^4 &, \nu'<\nu'_0\\
    \nu'^{1.1} &, \nu_0'<\nu'<\nu_\tr'\\
    \e^{-\frac{h \nu'}{k T}} &, \nu_\tr'<\nu'\,.
    \end{cases}
\end{equation}
We note that if the flux is integrated over the duration of the planar phase, the spectrum peaks at $\nu_0'$.

\begin{figure*}
    \centering
    \includegraphics[width = 0.49 \textwidth]{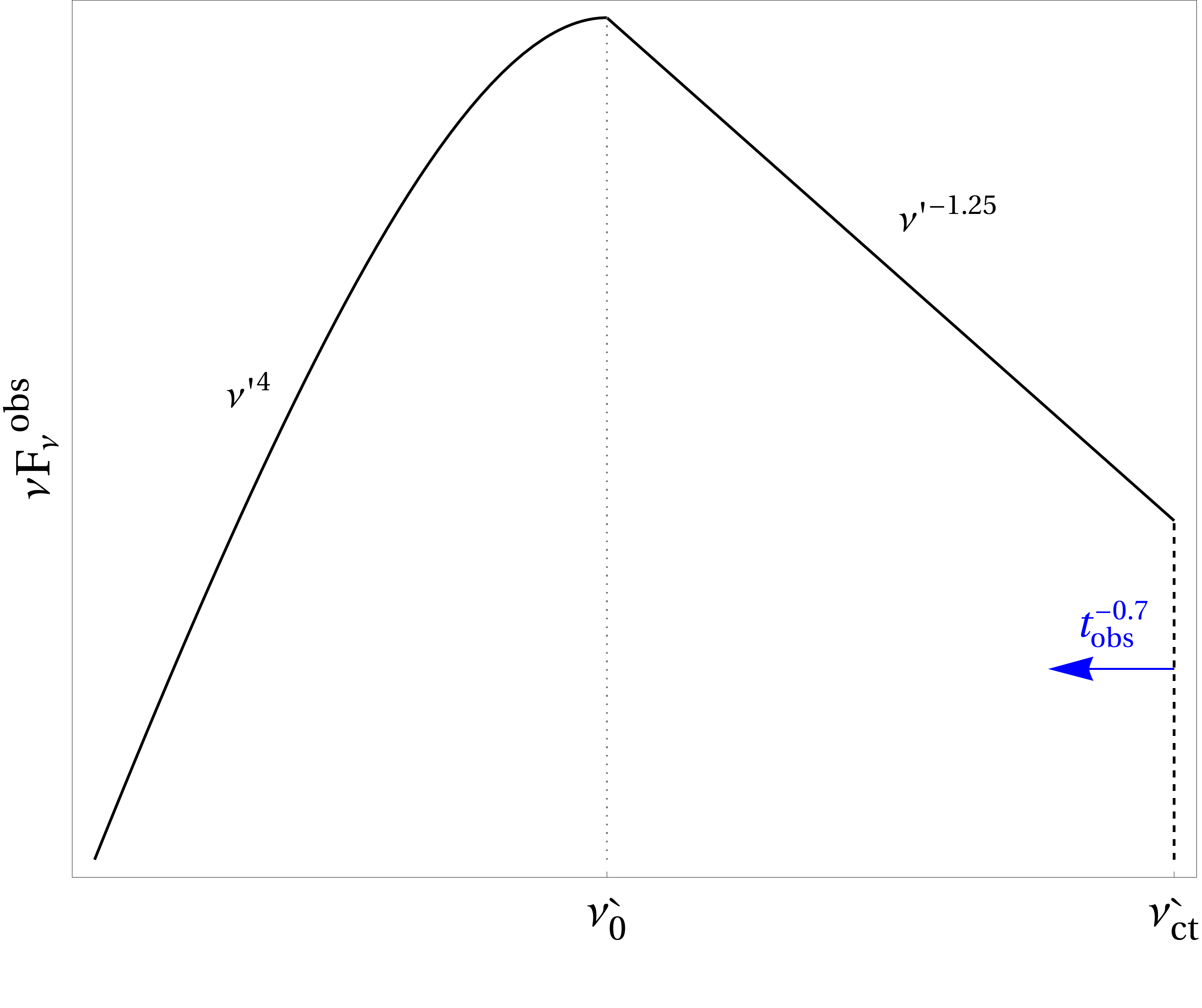}
    \includegraphics[width = 0.49 \textwidth]{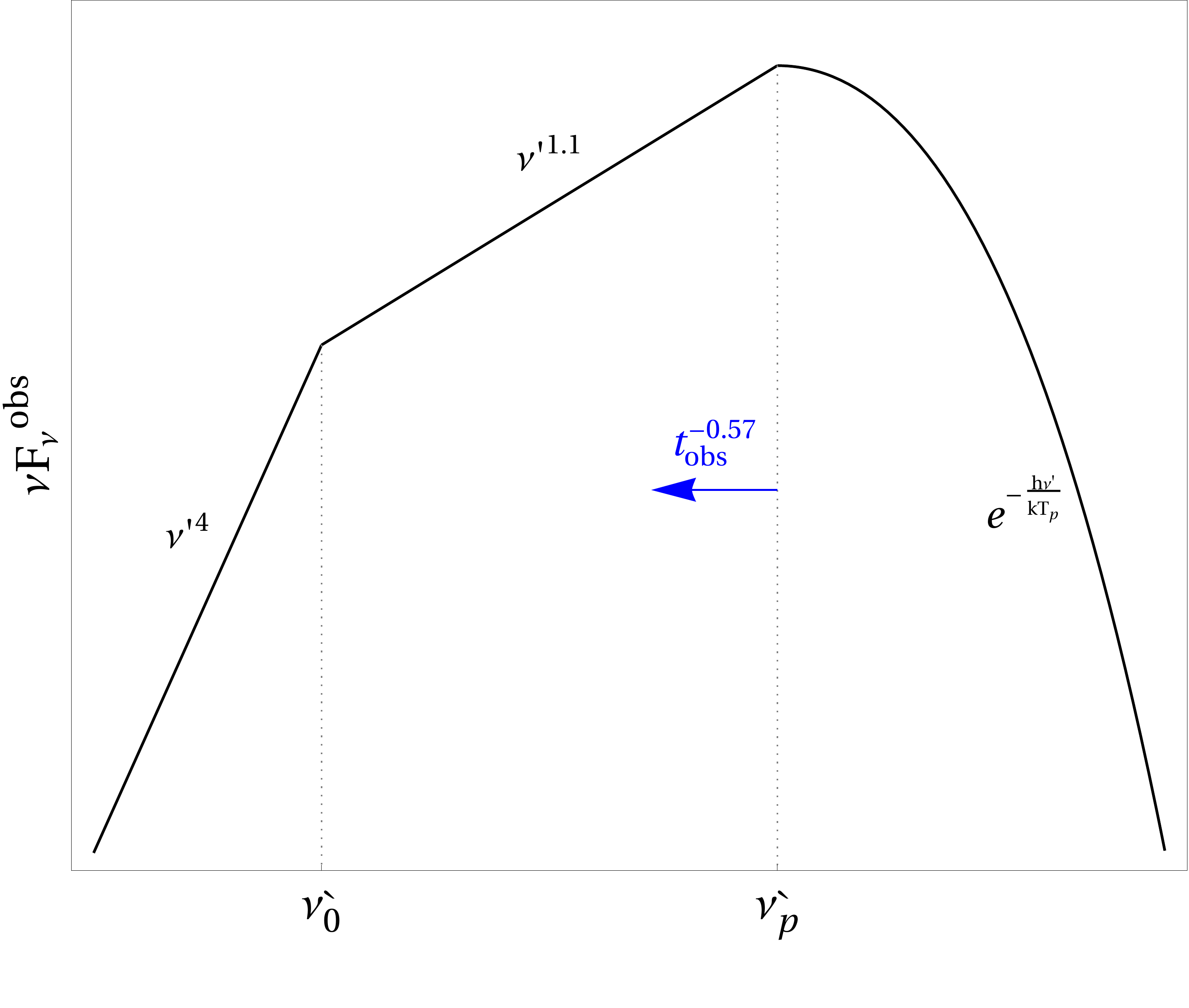}
    \caption{The instantaneous observed spectrum of the breakout pulse, at $t_0^\obs<t_\obs<t_\s^\obs$. Left: the case of a hot planar phase; the spectrum peaks at the photon energy corresponding to the temperature of the unloaded breakout shell. Since effectively no shells located at $m>m_0$ can contribute to the observed luminosity, the spectrum follows a Wien spectral density below the peak. The high frequency flux above the peak originates in hot shells located at $m<m_0$, where the high frequency cutoff corresponds to the peak photon energy of the shell that stops contributing to the luminosity at $t_\obs$. Right: the case of a cold planar phase, at times $t_\obs>t_0^\obs,t_{\c,2}^\obs$; the spectrum peaks at a frequency $\nu'_\p$, which corresponds to the temperature ($T_\p$) of the shell entering its spherical phase at $t_\obs$, and shifts to lower frequencies with time. Below the peak the spectrum is governed by emission from shells located at $m_\tr\leq m\leq m_0$.}
    \label{f:spectrum}
\end{figure*}

If the unloaded breakout shell is hot throughout the planar phase, it governs the spectrum at and below the peak, located at $x_0^\obs \sim 6\,\theta_{\i,0} \gamma_{\i,0}$.
The fact that shells stop contributing to the luminosity at $t_\obs\sim R_*/(2c\gamma_\tr^2)$ introduces a high frequency cutoff that decreases with time as $\nu_\ct' \propto t_\obs^{-0.70}$. The shape of the spectrum in this case at $t_0^\obs<t_\obs<t_\s^\obs$ is
\begin{equation}\label{eq:planar_spectrum_hot}
    \nu F_\nu^\obs|_\text{hot} \propto \begin{cases}  
    \nu'^4 &, \nu'<\nu_0'\\
    \nu'^{-1.25} &, \nu_0'<\nu'<\nu_\text{ct}'\\
    \e^{-\frac{h \nu'}{k T}} &, \nu_\ct'<\nu'\,.
    \end{cases}
\end{equation}
Both expressions for $\nu F_\nu^\obs$ above are written for $t_0^\obs<t_\obs<t_\s^\obs$. Schematic plots of the instantaneous (not time integrated) spectral densities are shown in Figure \ref{f:spectrum}.

At the end of the planar phase, the observed signal is dominated by high latitude emission from the unloaded breakout shell (equations \ref{eq:Lobs_m}-\ref{eq:nu_m}), until emission from spherically expanding shells becomes dominant.
In principle, after the pair unloaded breakout shell has become transparent, a fraction of the energy is able to escape from deeper shells given many multiplications of the dynamical time prior to the onset of the spherical phase. This is expected to introduce logarithmic corrections to the luminosity and temperature \citep{Faran2019b}. A handle on the importance of the logarithmic corrections is given by $\log\big(\frac{t_\s}{t_0}\big) = 6\, \big(\frac{M_\ej}{5M_\odot}\big)^{1/4}\big(\frac{R_*}{5 R_\odot}\big)^{-1/2}$, and can be non-negligible if $1\ll \log(t_\s/t_0)$. Nevertheless, we do not address these corrections in our analysis.
\subsection{Spherical Phase Emission}
\subsubsection{Relativistic Phase}
During spherical expansion, $\tau_\tot \propto t'^{-2}$ and radiation escapes from shells satisfying $m_0<m$. At this point, the envelope does not contain any more pairs since they effectively vanished at $t_0^\obs$. The coordinate that satisfies $\tau_\T = 1$ is
\begin{equation}\label{eq:m_tr_sph_hot}
    m_\tr = m_0\bigg(\frac{t'}{R_*/c}\bigg)^2 ~.
\end{equation}
Shells that become transparent at $t_\s^\obs$ are still accelerating if $\gamma_{\i,0}\gtrsim 3$.
In this case, the transition to the cold spherical phase occurs when $m_\tr$ intersects $m_\c$, which now decreases as $m_\c \propto t'^{-8.25}$.
The lab frame time at which $m_\c = m_\tr$ is
\begin{equation}\label{eq:tc_sph_lab}
    t'_{\c,\s}  \simeq 0.8\,\t'_\s (n_0 \sigma_\T R_*)^{-0.17}\gamma_{\i,0}^{0.80}\,,
\end{equation}
and the observed time is
\begin{equation}\label{eq:tc_s_obs}
    t_{\c,\s}^\obs  \sim \t_\s^\obs (n_0 \sigma_\T R_*)^{0.31}\gamma_{\i,0}^{-1.46}\,.
\end{equation}
The hot spherical phase, if it exists, is extremely short. Therefore, after the high-latitude emission phase ends, radiation originates from shells that already cooled during the spherical or planar phase.

If the spherical phase stars cold, all shells have cooled during the planar phase, and the luminosity is
\begin{equation}\label{eq:L_sph_cold}
\begin{split}
    L^\obs|_\text{cold} &= \frac{\gamma_{\f,\tr} E_\tr}{t_\obs} \simeq \frac{\gamma_{\i,\tr}^{1+\sqrt{3}} E_{\i,\tr} \Big(\frac{t_\s}{t_\i}\Big)^{-1/3}\Big(\frac{t}{t_\s}\Big)^{-1}}{t_\obs} = \\
    &=10\,\frac{m_\p c^2}{R_*/c}\frac{R_*^2}{r_\e^2}\gamma_{\i,0}^{1+3(1+\sqrt{3})}(n_0\sigma_\T R_*)^{-0.21}\bigg(\frac{t_\obs}{t_\s^\obs}\bigg)^{-1.07}\,,
\end{split}
\end{equation}
Since the number of photons in each optically thick shell remains constant, the rest frame temperature of fluid elements evolves adiabatically as $\theta \propto t'^{-1}\propto t^{-1}$. The observed temperature is therefore
\begin{equation}\label{eq:T_sph_cold}
\begin{split}
    \theta^\obs|_\text{cold} &= \gamma_{\f,\tr}\theta_\tr \simeq \gamma_{\i,\tr}^{1+\sqrt{3}} \theta_{\i,\tr} \bigg(\frac{t_\s}{t_\i}\bigg)^{-1/3}\bigg(\frac{t}{t_\s}\bigg)^{-1} = \\
    &=1.75\,\gamma_{\i,0}^{2.75}\theta_{\i,0}(n_0\sigma_\T R_*)^{-0.49}\bigg(\frac{t_\obs}{t_\s^\obs}\bigg)^{-0.64}\,,
\end{split}
\end{equation}
If the spherical phase starts hot, the luminosity shell first probes regions that have cooled during the spherical phase, for which $\gamma_\f$ is given by Eq \eqref{eq:gamma_f_sph}. However, as it propagates deeper into the stellar ejecta, photons start leaking out of shells that have already cooled during the planar phase, for which $\gamma_\f$ is given by Eq \eqref{eq:gamma_f_planar}. We denote the time when $m_\tr = m_\c(t_\s)$ as $t^\obs_{\c,\pl}$:
\begin{equation}
    t^\obs_{\c,\pl} = 0.02\, \frac{R_*}{c}(n_0 \sigma_\T R_*)^{-1.48} \gamma_{\i,0}^{5.0}\,.
\end{equation}
The observed luminosity and temperature in the hot case obey
\begin{equation}\label{eq:L_sph_hot}
\begin{split}
    L^\obs|_\text{hot} &= \frac{\gamma_{\f,\tr} E_\tr}{t_\obs} \simeq \frac{m_\p c^2}{R_*/c}\frac{R_*^2}{r_\e^2}\gamma_{\i,0}^{7.20}(n_0\sigma_\T R_*)^{-0.89}\\
    &\times\begin{cases}
        \Big(\frac{t_\obs}{t_{\c,\s}^\obs}\Big)^{-1.01} & , t_{\c,\s}^\obs<t^\obs<t^\obs_{\c,\pl}\\
       \Big(\frac{t^\obs_{\c,\pl}}{t_{\c,\s}^\obs}\Big)^{-1.01} \Big(\frac{t_\obs}{t^\obs_{\c,\pl}}\Big)^{-1.07} & , t^\obs_{\c,\pl}<t^\obs
    \end{cases}
\end{split}
\end{equation}
and
\begin{equation}\label{eq:T_sph_hot}
\begin{split}
    &\theta^\obs|_\text{hot} = \gamma_{\f,\tr}\theta_\tr = 0.8\,\gamma_{\i,0}^{1.30}\theta_{\i,0} (n_0 \sigma_\T R_*)^{-0.27}\\
    \times
    &\begin{cases}
    \Big(\frac{t_\obs}{t^\obs_{\c,\s}}\Big)^{-0.71}~, &t^\obs_{\c,\s}<t^\obs<t^\obs_{\c,\pl}\\
    \Big(\frac{t^\obs_{\c,\pl}}{t^\obs_{\c,\s}}\Big)^{-0.71}\Big(\frac{t_\obs}{t^\obs_{\c,\pl}}\Big)^{-0.64}~, &t^\obs_{\c,\pl}<t^\obs<t^\obs_{z_{+,\i}}\,,
    \end{cases}
\end{split}
\end{equation}
where $\gamma_\f$ is given by Eq \eqref{eq:gamma_f_sph} in the first line, and by Eq \eqref{eq:gamma_f_planar} in the second line, and $t^\obs_{z_{+,\i}}$ is the observed time at which the luminosity shell reaches regions that were not pair loaded at $t_\i$. In shells that were never rich in pairs, the temperature decreases rapidly with $m$ until reaching thermal equilibrium. An estimate for the mass of the shell that satisfies $z_+ = 1$ at $t_\i$ can be obtained using Eq \eqref{eq:x_pos_DC}, since in the internal parts of the envelope photon production is dominated by DC:
\begin{equation}\label{eq:m_x_1}
    m(z_+=1) = \frac{\pi}{32 \alpha}\frac{1}{\theta_\i^2 \Lambda_\i g_\DC(\theta_\i)} \simeq 400\, m_0 \,.
\end{equation}
This coordinate (with respect to $m_0$) is insensitive to the density of the envelope and to the shock Lorentz factor. Assuming that this shell cooled during the planar phase, it is exposed at
\begin{equation}\label{eq:t_z1_rel}
    t^\obs_{z_+,\i}|_\text{rel} \simeq 800\,\frac{R_*}{c} \gamma_{\i,0}^{-5.46}\,,
\end{equation}
given that this shell is relativistic.

\subsection{Newtonian Phase}
Since the Lorentz factor of the flow decreases with mass, at some point the observed dynamics become Newtonian. If $\gamma_{\i,0}\beta_{\i,0}>1$, this transition occurs during the spherical phase. The shell that satisfies $\gamma_\f\beta_\f\sim1$ is located at $m_\NW= m_0 \gamma_{\i,0}^{5.75}$. The transition to the Newtonian phase happens when the luminosity shell reaches $m_\NW$, at
\begin{equation}
    t^\obs_\NW \simeq 0.13\, \frac{R_*}{c}\gamma_{\i,0}^{2.87}\,.
\end{equation}
In the above expressions for $m_\NW$ and $t_\NW$ we assumed that the ultra-relativistic dynamics hold down to the point where $\gamma_\f\beta_\f \sim 1$. However, the ultra-relativistic solution is not expected to be accurate at Lorentz factors lower than $\sim 2$, and therefore the above expressions are only approximate. Newtonian evolution implies that the luminosity shell satisfies $\tau_\T = c/v(m)$ and evolves as $m_\tr \propto t_\obs^{1.75}$.
If $m_\NW>m(z_+=1)$, then the luminosity and temperature at $t_\obs>t^\obs_\NW$ obey the Newtonian evolution studied in various previous works \citep[e.g.,][]{Nakar2010,Rabinak2011,Faran2019b}. If, however, $m_\NW<m(z_+=1)$, the initial temperature at $m_\NW<m<m(z_+=1)$ is regulated by pairs, and the observed temperature decreases like
\begin{equation}\label{eq:T_NW}
    \theta^\obs = \theta^\obs_\NW\bigg(\frac{t_\obs}{t_\NW^\obs}\bigg)^{-0.60}\,,
\end{equation}
where $\theta_\NW^\obs$ is the observed temperature at $t_\NW^\obs$ and $t_\obs = t' = t$. Since pairs do not exist in the envelope deeper than $\sim 400\,m_0$, this regime exist only if $\gamma_{\i,0}\lesssim 3$. The time when shells that were never pair loaded are exposed in this case is
\begin{equation}\label{eq:t_z1_NW}
    t^\obs_{\z_+,\i}|_\NW \simeq 4\, \frac{R_*}{c} \gamma_{\i,0}^{-0.41}\,.
\end{equation}
The luminosity evolves as $L\propto t_\obs^{-0.35}$ \citep[e.g.,][]{Nakar2010}, regardless of whether the probed shells used to be pair loaded or not.
\section{Applications}\label{sec:applications}
Given a powerlaw density index $n$ and a pre-explosion stellar radius, $R_*$, the problem depends only on the physical properties of the unloaded breakout shell: $\gamma_{\i,0}$ and $n_0$. We use \cite{Tan2001} to infer those values for given ejecta mass $M_\ej$ and the progenitor radius $R_*$ along with the explosion energy $E_\exp$, assuming $n=3$. The shock wave obeys Newtonian dynamics in the interiors of the star and accelerates to relativistic velocities at the outer envelope. In this case, the number density and velocity of the unloaded breakout shell satisfy
\begin{equation}\label{eq:n0_M_R}
    n_0 = 9\times 10^{15}\,\text{cm}^{-3}\, M_{\ej,5}^{1/4}R_5^{-3/2}\,,
\end{equation}
and
\begin{equation}\label{eq:gammabeta0_M_R}
    (\gamma\beta)_{\i,0} = 2\,  M_{\ej,5}^{-0.44}R_5^{-0.35}E_{53}^{0.62}A_v^{1.24}\,,
\end{equation}
where $M_x$ is the ejecta mass in units of $x \, M_\odot$, $R_x$ is the pre-explosion stellar radius in units of $x\,R_\odot$ and $E_x$ is defined by $E = 10^x$ erg. $A_v$ is a coefficient that depends on the propagation of the non relativistic shock wave in the internal parts of the stellar envelope (and hence on the internal density profile; see equation 4 in \citealt{Tan2001}), and is typically close to unity. We keep the scaling of $A_v$ since the observables are often sensitive to its value. We note that in the hydrodynamic solution leading to Eq \eqref{eq:gammabeta0_M_R} it was assumed that the gravitational binding energy of the star can be neglected.
Plugging equations \eqref{eq:n0_M_R}-\eqref{eq:gammabeta0_M_R} into equations \eqref{eq:t_s_obs}, \eqref{eq:tc_obs}, \eqref{eq:Lobs_planar_cold}, \eqref{eq:Tobs_planar_cold} and \eqref{eq:t_z1_NW}, we find the observed properties of the emission in the case where the unloaded breakout shell cools before the spherical phase:
\begin{subequations}
\begin{alignat}{7}
    t_\c^\obs|_\cold &\simeq \frac{5 d_{\c,\i}/c}{2 \gamma_{\f,\c}^2} \simeq 10^{-5}\, \text{s} \, M_5^{-0.89}R_5^{0.99}E_{52.6}^{0.89}A_v^{1.78}\\
    t_\s^\obs|_\cold  &\simeq \frac{R_*/c}{2 \gamma_{\f,0}^2} \simeq 1\, \text{s} \, M_5^{2.42}R_5^{2.93}E_{52.6}^{-3.39}A_v^{-6.77}\\
    t_\NW^\obs &\simeq 2 \, \text{s} \, M_5^{-1.27}R_5^{-0.02}E_{52.6}^{1.78}A_v^{3.56}\,,\\
    t_{z_{+,\i}}^\obs|_\NW &\simeq 50 \, \text{s} \, M_5^{0.18}R_5^{1.14}E_{52.6}^{-0.25}A_v^{-0.51}\\
    L_\p^\obs|_\cold &\simeq \frac{\gamma_{\f,\c} E_\c}{\frac{R_*/c}{2\gamma_{\f,\c}^2}}\nonumber \\& \simeq  10^{45}\, \text{erg s}^{-1}\, M_5^{-2.55}R_5^{-1.03}E_{52.6}^{3.56}A_v^{7.12}\\
    \theta_\p^\obs|_\cold &\simeq \gamma_{\f,\c}\, (0.6\times 50\, \text{keV}) \simeq\, 100\, \text{keV}\,,\\
    \theta_\text{p,int}^\obs|_\cold &\simeq \gamma_{\f,0} (0.6\times50\,\text{keV})\nonumber \\ &\simeq 60\, \text{keV}\, M_5^{-1.21}R_5^{-0.97} E_{52.6}^{1.69}A_v^{3.39}\,,
\end{alignat}
\end{subequations}
where $L_\p^\obs$ is the peak luminosity, $E_\c$ is the rest frame internal energy of $m_{\c,\tr}$ at $t_\c^\obs$,  $\theta_\p^\obs$ corresponds to the observed peak energy of the non-integrated spectrum, and $\theta^\obs_\text{p,int}$ corresponds to the observed peak energy of the time-integrated spectrum.
If the unloaded breakout shell remains hot throughout the planar phase, which requires an extreme explosion energy or a compact progenitor, the relevant observables are:
\begin{subequations}
\begin{alignat}{6}
    t_0^\obs|_\hot  &\simeq\,\frac{5 d_{i,0}/c}{2 \gamma_{\tr,0}^2} \simeq  10^{-6}\, \text{s} \, M_1^{0.64}R_1^{2.21}E_{53.5}^{-1.24}A_v^{-2.48}\\
    t_\s^\obs|_\hot & \simeq \frac{R_*/c}{2 \gamma_{\tr,0}^2} \simeq 10^{-3}\, \text{s} \, M_1^{0.89}R_1^{1.71}E_{53.5}^{-1.24}A_v^{-2.48}\\
    t_{z_{+,\i}}^\obs|_\text{rel} &\simeq 10^{-3} \, \text{s} \, M_1^{2.42}R_1^{2.93}E_{53.5}^{-3.39}A_v^{-6.78}\\
    L_\p^\obs|_\hot &\simeq \frac{\gamma_{\tr,0} E_0}{t_\s^\obs}\nonumber \\ &\simeq 6\times10^{49}\, \text{erg s}^{-1}\, M_1^{-1.77}R_1^{-0.41} E_{53.5}^{2.48}A_v^{4.96}\\
    \theta_\p^\obs|_\hot &\simeq \gamma_{\tr,0} (0.6\times50\,\text{keV})\\
    &\simeq 1\text{MeV}\, M_1^{-0.44}R_1^{-0.35}E_{53.5}^{0.62}A_v^{1.24}\\
    t_\NW^\obs &\simeq 700 \, \text{s} \, M_1^{-1.27}R_1^{-0.02}E_{53.5}^{1.78}A_v^{3.56}\,,
\end{alignat}
\end{subequations}
where $E_0$ is the rest frame internal energy of the unloaded breakout shell at $t_0^\obs$ and $L_\p^\obs$ is the luminosity at that time, which is also the peak luminosity.

\subsection{Closure relation for relativistic breakout}
Given that the planar phase $\gamma$-ray signal can be distinguished from that of the subsequent spherical evolution, the validity of our model for relativistic shock breakout can be observationally challenged. \cite{Nakar2012} showed that there exists a closure relation between $3$ observables (the duration of the planar phase pulse, the peak photon energy and the total pulse energy) and $2$ physical parameters of the system (the initial stellar radius and the initial Lorentz factor of the unloaded breakout shell). Assuming that the unloaded breakout shell reaches its final velocity before becoming transparent, the duration, temperature and total energy of the $\gamma$-ray pulse are
\begin{subequations}
\begin{alignat}{3}
    &\Delta t_\gamma = \frac{R_*/c}{2 \gamma_{\f,0}^2} \,,\label{eq:t_gamma_pl}\\
    &\theta_\p^\obs \simeq 0.6\, \gamma_{\f,0}\theta_{\i,0}\,,\label{eq:theta_peak}\\
    &E_\gamma^\obs \simeq 1.6\times 10^{45}\,\text{erg}\,R_5^2 \gamma_{\f,0}^{\frac{1+\sqrt{3}}{2}} \,,\label{eq:E_gamma_pl}
\end{alignat}
\end{subequations}
respectively. Equations \eqref{eq:t_gamma_pl}--\eqref{eq:E_gamma_pl} produce the following closure relation between the $3$ observed parameters
\begin{equation}\label{eq:closure_planar}
    \Delta t_\gamma =  9\, \bigg(\frac{E_\gamma^\obs}{10^{46}\,\text{erg}}\bigg)^{1/2} \bigg(\frac{\theta_\p^\obs}{0.1}\bigg)^{-\frac{9+\sqrt{3}}{4}}\,,
\end{equation}
where we used $\theta_{\i,0} = 0.15$ for the initial temperature of the unloaded breakout shell. We note that in the cold planar phase, $4\,\theta_0^\obs$ is the photon energy corresponding to the peak of the frequency integrated spectrum, $\nu F_\nu^\obs$. While the scaling laws in Eq \eqref{eq:closure_planar} agree with \cite{Nakar2012}, our somewhat different coefficient is based on a more detailed examination of the microphysics.



\subsection{low-luminosity GRBs}\label{sec:llGRBs}
Low-luminosity GRBs (llGRBs) are a subclass of long GRBs (LGRBs) that show several predominantly distinct observational features: their $\gamma$-ray isotropic equivalent energies are $3-4$ orders of magnitude lower than regular LGRBs, their spectra are softer and lack a high energy power-law tail, and their light curves are non-variable \citep[e.g.,][]{Woosley2006,Hjorth2012,Cano2017}. These differences suggest that another type of physical mechanism is responsible for $\gamma$-ray generation in llGRBs. Nevertheless, both classes are found to be associated with broad--line (BL) type Ic SNe, which points towards a common explosion mechanism, taking place in different environments. \cite{Bromberg2011} showed that many llGRBs have pulse durations that are inconsistent with the jet breakout time predicted by the Collapsar model. They conclude that the physical processes that cause the $\gamma$-ray emission in llGRBs must be different to those in LGRBs, and invoked relativistic shock breakout from the stellar material as one of the possible mechanism for generating the observed $\gamma$-ray emission. Despite being $10-100$ times more common than cosmological GRBs, the detection rate of llGRBs is low due to the small volume from which they can be detected. As a consequence, only a handful of llGRBs have been discovered. The small sample size of detected llGRBs makes it difficult to establish a theory that explains the observed properties of llGRBS and their physical connection to LGRBs and BL type Ic SNe. Therefore, it has not yet been determined whether the $\gamma$-ray emission in llGRBs arises from the same mechanism as LGRBs or whether they originate from relativistic shock breakout \citep{Kulkarni1998,Tan2001,Waxman2007,Bromberg2011,Nakar2012,Matzner2013,Duran2015}. 
The closure relation for relativistic shock breakout can serve as a test to the validity of our model for these events. We apply Eq \eqref{eq:closure_planar} to a sample of seven llGRBs compiled from the literature. For this analysis we assume that the $\gamma$-ray burst duration corresponds to the duration of the planar phase, while if $\gamma$-rays are also emitted during the spherical phase, Eq \eqref{eq:closure_planar} underestimates the observed pulse duration. The results are summarized in Table \ref{t:llGRBS}.

\emph{GRB980425/SN1998bw}: This event served as the first evidence for the association of llGRBs with BL type Ic SNe \citep{Galama1998}. Its isotropic equivalent energy of $E_\gamma^\text{iso} \sim 10^{48}\,\text{erg}$ and ejecta energy of $2-5\times 10^{52}\text{erg}$ \citep{Iwamoto1998} hinted at the existence of a relativistic shock wave \citep{Kulkarni1998,Tan2001}. 
Its observed duration $T_{90,\gamma}=23.3\pm1.4 \text{ s}$ and its  peak energy at $\epsilon_\p^\obs = 150-200$ keV, makes it a clear outlier of the Amati relation between $E_\iso$ and $\epsilon_\p^\obs$ \citep{Amati2006,Galama1998, Bloom1998,Kaneko2007}.

The reported $\gamma$-ray spectrum  in \cite{Kaneko2007} is integrated over a time interval of $\sim 5$ seconds. Since $t_\obs\,L_\gamma^\obs \propto t_\obs^{0.37}$, a spectrum integrated over the entire planar phase is expected to peak at $\nu_0'$ and to drop like $\nu'^{-0.64}$ at higher frequencies. 

For a Wien spectral distribution, the peak of the integrated spectrum is at $\h\nu =4k_\B T$, which implies that the observed temperature is $\sim 50$ keV.
Such a moderate temperature suggests that the shock is only mildly relativistic before breakout, and our closure relation is only approximate.
The closure relation predicts a duration of $\sim 93$ s using $\theta_{\i,0} \sim 0.15$, which is in agreement with the observed $T_{90,\gamma}$, reinforcing a shock breakout origin.
We estimate that $\gamma\beta_{\f,0}\sim 1$ according to Eq \eqref{eq:theta_peak}, so that the initial Lorentz factor was $\gamma\beta_{\i,0}\sim 0.8$.
The inferred progenitor radius is $\sim 3\times 10^{12}$ cm (Eq \ref{eq:t_gamma_pl}), and the explosion energy is $\sim 2\times 10^{53}$ erg. Since $\gamma_{\i,0}<3$, the unloaded breakout shell is cold when it becomes optically thin, and the spectrum is expected to obey Eq \eqref{eq:planar_spectrum_cold}. 

\emph{GRB 031203/SN2003lw}: The burst emitted a total $\gamma$-ray energy of $E{\gamma,\iso} \sim 8\times10^{49}\,\text{erg}$, over a timescale of $T_{90,\gamma} = 36$ s \citep{Sazonov2004,Kaneko2007,Cano2017}. Since the peak of the $\gamma$-ray distribution is not clearly inferred from the INTEGRAL observations, we adopt the maximal observed photon energy as a lower limit for the peak energy of the emission. The spectra presented in \cite{Sazonov2004} and \cite{Kaneko2007} are integrated over a duration of 22 s and 46 s, respectively, and from both a lower limit of $\epsilon_\p^\obs\gtrsim 300$ keV can be put on the observed photon energy, implying a temperature higher than $\sim 75$ keV. Since the integration time is similar to the duration of the burst, the peak of the spectrum is expected to correspond to the temperature of the unloaded breakout shell. The closure relation predicts $\Delta t_\gamma \sim 280$ s for the observed parameters, which is of the same order of magnitude as the observed pulse duration. The closure relation is exactly satisfied for a peak energy of $\sim 650$ keV.

\emph{GRB 060218/SN2006aj}: GRB 060218 displayed several unique features that are difficult to explain by a standard GRB theory: a long lasting $\gamma$-ray emission, soft and bright X-ray afterglow and a prominent thermal component in the X-ray and optical bands at early times \citep{Campana2006,Kaneko2007}.
Relativistic shock breakout was deployed by several authors to explain the observed emission \citep[e.g.,][for a discussion on alternative models see \citealt{Irwin2016}.]{Campana2006,Waxman2007,Duran2015}.
The peak energy of the spectrum is very low: $\epsilon_\p^\obs\sim 40$ keV, corresponding to a temperature of $10$ keV. This temperature is too low to suggest the presence of pairs in the unloaded breakout shell, and therefore the closure relation is not expected to be satisfied. It is however possible that GRB 060218 was a result of a sub relativistic shock breakout.

\emph{GRB 080517}:
The $\gamma$-ray energy released in this burst was $E_{\gamma,\iso} = (1.03\pm 0.21)\times 10^{49}$ erg within $T_{90,\gamma} = 65\pm 27$ s \citep{Stanway2015}. Only a lower limit can be placed on the a peak photon energy, $\epsilon_\p^\obs >55$ keV, which does not allow us to test the validity of the shock breakout model through the closure relation. 

\emph{GRB 100316D/SN2010bh}: This event resembles GRB 060218 in its long duration ($T_{90,\gamma}>1300$s) and low peak energy, $\epsilon_\p^\obs = 30-40$ keV \citep{Starling2011}. Due to the low peak energy, this event was most likely not a result of a relativistic shock breakout, although we again cannot rule out a shock breakout of a sub-relativistic nature.

\emph{GRB 171205A/SN2017iuk}: The peak photon energy detected in this event was relatively high, with $\epsilon_\p^\obs = 125^{+141}_{-37}$ keV \citep{Delia2018,Suzuki2019}, making it a good candidate for originating from a relativistic shock breakout. The total $\gamma$-ray energy measured was $E_{\gamma,\iso} = 2.18^{+0.63}_{-0.50}\times 10^{49}$ and was spread over $T_{90,\gamma} = 190.5\pm 33.9$ s. The spectrum shown in \cite{Delia2018} was integrated in time over $\sim 144$ s, similar to the burst duration, and therefore the peak photon energy corresponds to the temperature of the unloaded breakout shell. For these parameters, Eq \eqref{eq:closure_planar} predicts a duration of $\sim 10^3$ s, consistent with the observed pulse duration, supporting a shock breakout origin for this event.

\emph{GRB 190829A/SN2019oyw}: This GRB event had two peaks, separated by a quiescent phase - the first pulse was short and hard and is an outlier of the Amati relation, with $\epsilon_\p^\obs = 120^{+110}_{-37}$ keV, $E_{\gamma,\iso} = 3.2\times 10^{49}$ erg and $T_{90,\gamma} = 6$ s. The second burst was soft, more energetic and satisfies the Amati relation, with $\epsilon_\p^\obs = 10.9$ keV, $E_{\gamma,\iso} = 1.9\times 10^{50}$ erg, and $T_{90,\gamma} = 10$ s \citep{Lesage2019,Chand2020}. We treat the first peak as a candidate for a relativistic shock breakout, but find that the closure relation is not satisfied for a peak energy of $\epsilon_\p^\obs = 120$ keV, which predicts a burst duration of $800$ seconds. \cite{Chand2020} concluded that this energy satisfies the closure relation in \cite{Nakar2012}, since they wrongly treated $\epsilon_\p^\obs$ as the photon temperature, while $\epsilon_\p^\obs\sim 4 kT_\p$.
We note that since the closure relation is sensitive to the photon energy, taking the upper limit for $\epsilon_\p^\obs\simeq 330$ keV brings the predicted pulse duration to $\sim 50$ s, closer to the observed one. The association of this burst with a relativistic shock breakout is therefore inconclusive.

To summarize, within the sample of seven known llGRBs, we find that relativistic shock breakout as the origin of the observed emission is consistent with 3 of them (GRB980425, GRB031203 and GRB 171205A), is inconclusive for two (GRB 080517 and GRB 190829A) and is inconsistent for the rest (GRB060218 and GRB100316D), which could result from sub-relativistic shock breakouts.

\begin{table*}
  \caption{llGRBS closure relation values}\label{t:llGRBS}
    \centering 
  \begin{threeparttable}
\noindent\makebox[\textwidth]{  \hskip-2.0cm\begin{tabular}{ccccccll}
    \midrule
    GRB  & $T_{90,\gamma}$ [s] & $E_{\gamma,\iso}$ [erg] & $\epsilon_\p^\obs$ [keV]\tnote{1}& $\Delta t_\gamma$ [s]\tnote{2} & $\gamma\beta_\f$ & SBO\tnote{3}& Reference\\
    \midrule\midrule
    980425& $23.3\pm1.4$ & $10^{48}$ & $150-200$& $90$ &$\sim 1.2$&V& \cite{Galama1998,Kaneko2007} \\
    031203& $36$ & $8\cdot10^{49}$ & $\geq 300$& $280$ &$>2$&V& \cite{Kaneko2007,Cano2017}  \\
    060218& $2100$ & $(2-4)\cdot 10^{49}$ & $40$& $3\cdot10^4$ &$<1$ &X& \cite{Campana2006,Kaneko2007} \\
    080517& $65\pm 27$ & $10^{49}$ & $>55$& $10^4$ &$<1$&?& \cite{Stanway2015} \\
    100316D& $\geq1300$ & $\geq6\cdot 10^{49}$ & $30-40$& $5\cdot 10^4$ &$<1$&X&\cite{Starling2011} \\
    171205A& $190^{141}_{-37}$ & $(1-2)\cdot10^{49}$ & $125$& $10^3$ &$\sim1$& V&\cite{Delia2018,Suzuki2019} \\
    190829A \tnote{4}& $6$ & $3.2\cdot10^{49}$ & $120^{+112}_{-37}$& $2000$ &$\sim 1$&?& \cite{Lesage2019,Chand2020} \\
    190829A \tnote{5}& $10$ & $1.9\cdot10^{50}$ & $10.9$& $3\cdot 10^6$ &$\ll 1$&X& \cite{Lesage2019,Chand2020} \\
    \midrule\midrule
     \end{tabular}
     }
     \begin{tablenotes}
     \item[1] Peak energy of $\nu F_\nu^\obs$.
      \item[2] Pulse duration according to Eq \eqref{eq:closure_planar}
      \item[3] Consistency with the shock breakout model according to Eq \eqref{eq:closure_planar}.
        \item[4] Properties of the first burst, integrated over $4$ s.
        \item[5] Properties of the second burst, integrated over $4$ s.
     \end{tablenotes}
 \end{threeparttable}
\end{table*}
\subsection{White Dwarfs}\label{sec:WD}
Type Ia SN, which are explosions in a WD, carry a typical energy of $E\sim 10^{51}$ erg, ejecting a total mass of $M_\ej \sim 1.4\,M_\odot$ 
\citep{Woosley1986}. Taking a typical radius of $R_*\sim 2\times10^8$ cm, this explosion would yield a mildly relativistic shock wave with $(\gamma\beta)_{\i,0}\sim 1-2$. Indeed, velocities as high as $40,000$ km s$^{-1}$ were measured in e.g., SN2009ig \citep{Foley2012}, SN1991T and SN1990N \citep{Jeffery1992}. Given these parameters, our model predicts that a total energy of $\sim 10^{40}$ erg to be carried away by $\sim 1$ MeV photons during the planar phase; the peak luminosity of the $\gamma$-ray flare to be $\sim 10^{44}$ erg s$^{-1}$; and it will be spread over a duration of $\sim 10^{-5}$ s.
Other types of explosions (such as type .Ia and type Iax SNe) eject only a fraction of the WD. Known examples are SN2005E with $M_\ej \simeq 0.3 M_\odot$ and $E\simeq 4\times 10^{50}$ erg \citep{Perets2010}, SN2010X with $M_\ej \simeq 0.16 M_\odot$ and $E\simeq 1.7\times 10^{50}$ erg \citep{Kasliwal2010}, and SN2002bj with $M_\ej \simeq 0.2 M_\odot$ and $E\sim 10^{50}$ erg \citep{Poznanski2010}. The observed characteristics of the $\gamma$-ray pulse are similar to the case in which the entire envelope is ejected.

\subsection{Neutron stars}
Neutron stars, as far as we know, do not have a generic mechanism of explosion. Here, we nevertheless predict the outcome of a sudden release of a fiducial energy of $4\times 10^{51}$erg. In such an explosion, the gravitational binding energy cannot be neglected relative to the energy of the explosion, therefore the estimate for $(\gamma\beta)_{\i,0}$ in Eq \eqref{eq:gammabeta0_M_R} is not valid.
Instead, as a test case we use the numerical results presented in section 2.6 of \cite{Tan2001} to estimate $(\gamma\beta)_{\i,0}$. For an explosion in which $0.017M_\odot$ carry away $4\times 10^{51}\text{erg}$, the most external $10^{-5}\,M_\odot$ move at $(\gamma\beta)_\f\sim 6$, corresponding to $(\gamma\beta)_\i \sim 1.5$, (using Eq \ref{eq:gamma_f_planar}). For a NS with $R_* = 10^6 \text{ cm}$, we find $m_0 \sim 10^{-20}M_\odot$ and $n_0\sim 8\times 10^{22}\text{ cm}^{-3}$. Hence, the shock breaks out of a very thin layer in the atmosphere of the NS, where the pre-shocked material is cold ($T<10^6$ K) and neutron free. The atmospheric composition of NSs is uncertain. However, due to immense gravitational fields at their surface, heavy elements tend to sink into deeper layers, which motivates us to assume a pure hydrogen composition for simplicity (however, we note that accretion or strong magnetic fields can enrich the atmosphere in heavy elements). If the atmosphere is composed of heavier elements, other opacity sources like bound-bound and bound-free absorption should be taken into account, and the free-free emission rate would increase by a factor $Z^2$ where $Z$ is the atomic number, affecting the post shock temperature and pair density. The initial Lorentz factor of the unloaded breakout shell for the above parameters, assuming that at the edge of the star gravity no longer affects the dynamics, is $\gamma_{\i,0}\sim 700$. Due to the high velocities, the timescales are extremely short, and the planar phase only lasts $\sim 10^{-11}$ s. The observed luminosity is $\sim 10^{51}$ erg s$^{-1}$ and the total observed energy released is $\sim 10^{41}$ erg. The peak observed photon energy is $\sim 0.2$ GeV, and as a result the $\gamma$-ray flash will continue beyond the planar phase. The entire $\gamma$-ray flash could be observed by the \textit{Fermi} satellite, which can detect photons in the range $10$ keV--$300$ GeV.

Given the high densities and post shock temperature, one needs to verify that the propagation of a shock wave in the NS envelope is not impeded by production and then the escape of neutrinos. If cooling by neutrinos is efficient and the medium is optically thin to them, the shock may die out before breaking out of the star. The main neutrino production mechanisms we consider are electron capture $p+e^- \rightarrow n + \nu$, electron-neutrino bremsstrahlung $e^-+e^- \rightarrow e^-+e^-+\nu+\bar{\nu}$, and electron-positron pair annihilation $e^-+e^+ \rightarrow \nu + \bar{\nu}$. The cross section for electron-proton (photon) bremsstrahlung is several orders of magnitude larger than those of electron capture and electron-neutrino bremsstrahlung. Assuming that the energy carried by each photon and neutrino produced by those processes is approximately the same, the energy lost to neutrinos is negligible. 
Now, if electron-positron pairs exist in the medium, pair annihilation into a $\nu \bar{\nu}$ pair competes with pair annihilation into photons, but becomes more important only when $\gamma_\e\gtrsim 10^5$, where $\gamma_\e$ is the thermal Lorentz factor of the electrons. However, the electrons do not reach this temperature if the shock Lorentz factor is less than $10^5$.
The above arguments contradict the claim made by \cite{Yalinewich2017}, that shock waves cannot accelerate to high Lorentz factors in the atmospheres of NSs due to vigorous neutrino production and escape.

\section{Summary and Discussion}\label{sec:summary}
We calculated the observed properties of the emission following the emergence of a relativistic shock wave from a stellar edge. The essential principles of our steady state model are as follows: (1) equilibrium between electron-positron pair production and annihilation, (2) photon production by free-free or double-Compton mechanisms, neglecting absorption, (3) the downstream radiation field attains a Wien spectral distribution. The first and last conditions imply that the plasma is in a state of Wien equilibrium. Given a density power-law index $n$, our model depends only on three physical parameters: the pre-shock proton number density of the unloaded breakout shell $n_0$, its initial Lorentz factor $\gamma_{\i,0}$, and the stellar radius $R_*$ (or alternatively $R_*$, $M_\ej$ and $E_\exp$). Given these parameters, we calculate the post-shock state of the envelope and analytically compute its temperature and pair density profiles as a function of time. We then deduce the luminosity and spectrum of the escaping emission as the envelope becomes transparent. The main findings of our work are listed below.

\begin{enumerate}[wide, labelwidth=!, labelindent=0pt]
    \item We show that the initial temperature of the plasma, immediately after shock passage and before substantial expansion of the ejecta, depends very weakly on both the density and the shock Lorentz factor. We also show that the initial temperature profile in pair loaded shells is relatively shallow, where $\theta_\i \propto m^{-0.07}$, ranging typically between $\sim 0.2$ in the breakout shell to $\sim 0.07$ deeper inside the envelope.
    \item At the beginning of the planar phase the envelope is pair loaded and opaque. Adiabatic cooling quickly decreases the optical depth through pair annihilation, and photons escape freely from the shell satisfying $\tau_\tot = 1$. Owing to the exponential dependence of pair density on the temperature, shells become optically thin after roughly two multiplications of their dynamical time, resulting in only a modest decrease in the initial rest frame temperature prior to the release of energy.
    \item The most internal shell that contributes to the planar phase emission is the unloaded breakout shell, whose pair unloaded optical depth is unity, and hence becomes transparent once all pairs disappear from the envelope. This shell dominates the time integrated energy released during the planar phase, since the shock deposits most of its energy in massive shells.
    \item Due to relativistic beaming and light travel time effects, high latitude emission coming from external shells is mixed with low latitude emission of more internal regions, where each shell releases its energy on a timescale of order $\frac{R_*/c}{2\gamma_\tr^2}$. During the very early planar phase, the luminosity \textit{increases} as $L_\obs\propto t_\obs^{0.51}$, while the temperature decreases as $T_\obs\propto t_\obs^{-0.41}$.
    \item The planar phase has a very short observed duration of $t_\s^\obs = \frac{R_*/c}{2 \gamma_{\tr,0}^2}$, typically less than a few seconds. Planar phase emission manifests as a flash of energetic $\gamma$-ray photons, with a time integrated $\nu F_\nu ^\obs$ spectrum peaking at $\sim 200\,\gamma_{\tr,0}$ keV, dominated by the energy deposited in the unloaded breakout shell. These characteristics form a closure relation between the three observables: $\Delta t_\gamma$, $E_\gamma^\obs$ and $\theta_\p^\obs$, and provides a test for the shock breakout model.
    \item If $\gamma_{\i,0}\lesssim 1.2$, the unloaded breakout shell exhausts its thermal energy and reaches its terminal Lorentz factor before the spherical phase and before becoming optically thin. If, however, the shock wave propagates in the external density profile of an especially compact star (e.g., a NS), the unloaded breakout shell may obtain higher Lorentz factors, and the acceleration of the envelope continues during the spherical phase if $\gamma_{\i,0}\gtrsim 3$. When $1.2\lesssim\gamma_{\i,0}\lesssim 3$, shells that are internal to the unloaded breakout shell continue to accelerate and reach $\gamma_\f$ between the time the latter is exposed and the beginning of the spherical phase.
    \item The initial existence of pairs at shock breakout continues to affect the temperature of the plasma and the observed emission even after their annihilation. Since we obtain the pair density profile in the envelope, we are able to predict the time at which pairs no longer affect the observed temperature, denoted $t^\obs_{z_+,\i}$. Afterwards, the temperature decrease quickly, while the luminosity is not affected.
\end{enumerate}

The observational signature of a relativistic shock breakout is different from its Newtonian counterpart. Since relativistic shock waves enhance the optical depth through pair production, they are able to propagate further along the stellar edge, and accelerate to increasingly high Lorentz factors before breakout. In addition, shocked fluid elements can accelerate significantly after shock breakout while converting thermal energy to bulk kinetic energy. On the other hand, in Newtonian flows the post-shock thermal and kinetic energy contents are comparable, and the velocity can only increase by a factor of $\sim \sqrt{2}$. In addition, the high Lorentz factor of the flow shortens the time scales over which photons arrive at the observer. The emission therefore appears as a short $\gamma$-ray pulse.

Several differences exist between this work and that of \cite{Nakar2012}. One major difference comes from their assumption that the plasma is initially heated by the shock to a temperature of $200$ keV, and releases its energy once $T\sim 50$ keV. This assumption means that prior to the release of internal energy, the rest frame time has increased by a factor of $\sim 60$, which results in more significant adiabatic losses. In contrast, we find that shells become transparent only after two multiplications of the dynamical time and compute the temperature of the emitted radiation more accurately, using the initial post-shock temperature profiles. The pair density profile, which is also computed in regions where double-Compton is dominant over free-free emission, allows us to better estimate  the time at which pairs stop affecting the observed spectrum. In addition, \cite{Nakar2012} limit their work to cases in which the acceleration of the unloaded breakout shell ends during the planar phase. We allow for arbitrarily high shock Lorenz factors, so that our model is adequate to describe shock breakout from compact objects such as NSs, in which spherical acceleration can take place.

We update the closure relation for relativistic shock breakout that was introduced by \cite{Nakar2012} and apply it to $\gamma$-ray observations from llGRBs. Among the seven events that we examine, three are consistent with our model for relativistic shock breakout and show evidence for relativistic ejecta. The rest of the cases are inconclusive, and cannot be confirmed nor ruled out as having a  shock breakout origin, since the temperature inferred from their $\gamma$-ray spectra is less than $50$ keV. It is possible that these bursts are nonetheless a result of sub-relativistic shock breakouts, which are expected to have a typical radiation temperature lower than $50$ keV.

We apply our model to shock breakout from WDs following a SN explosion. A typical type Ia SN explosion is expected to produce mildly relativistic ejecta, with $(\gamma\beta)_{\i,0}\sim 1-2$, and release a total energy of $10^{40}$ erg in $\sim 1$ MeV $\gamma$ ray photons. Indeed, some measurements of type Ia SN ejecta velocities of $v\sim40,000$ km s$^{-1}$ support this finding.


The only $\gamma$-ray emission detected from a binary neutron star merger, which was associated with GW170817, is thought to have been produced in a relativistic shock breakout \citep[e.g.,][]{Kasliwal2010,Gottlieb2018,Bromberg2018,Beloborodov2020,Lundman2021}. The first electromagnetic signal from the event arrived as a short flash of energetic $\gamma$-ray photons, 1.7 seconds post merger. A plausible source for this radiation is the breakout of a relativistic shock wave from the merger ejecta, driven by a highly pressurized bubble known as the `cocoon' \citep{Gottlieb2018}. However, the conditions in BNS merger outflows are very different from those at a stellar edge, and are expected to affect the observed signal. The heavy composition, rich in r-process elements, increases the opacity and the photon emissivity, which in turn highly affect the pair fraction. In addition, the shock wave in these systems is propagating into an ejecta that was previously shocked, and might be moving at relativistic velocities, so that the hydrodynamic solution applied in this work does not describe the BNS shocked ejecta. Therefore, some modifications should be made to our model in order to make it applicable to the outflows of BNS mergers, and will be implemented in future work.
\section*{}
The authors thank Ehud Nakar and Tsvi Piran for helpful discussions. This work was supported by an advanced ERC grant TReX and an ISF grant.
\appendix
\section{Wien Spectrum}\label{app:Wien}
\begin{figure*}
\centering
\includegraphics[width=0.5\columnwidth]{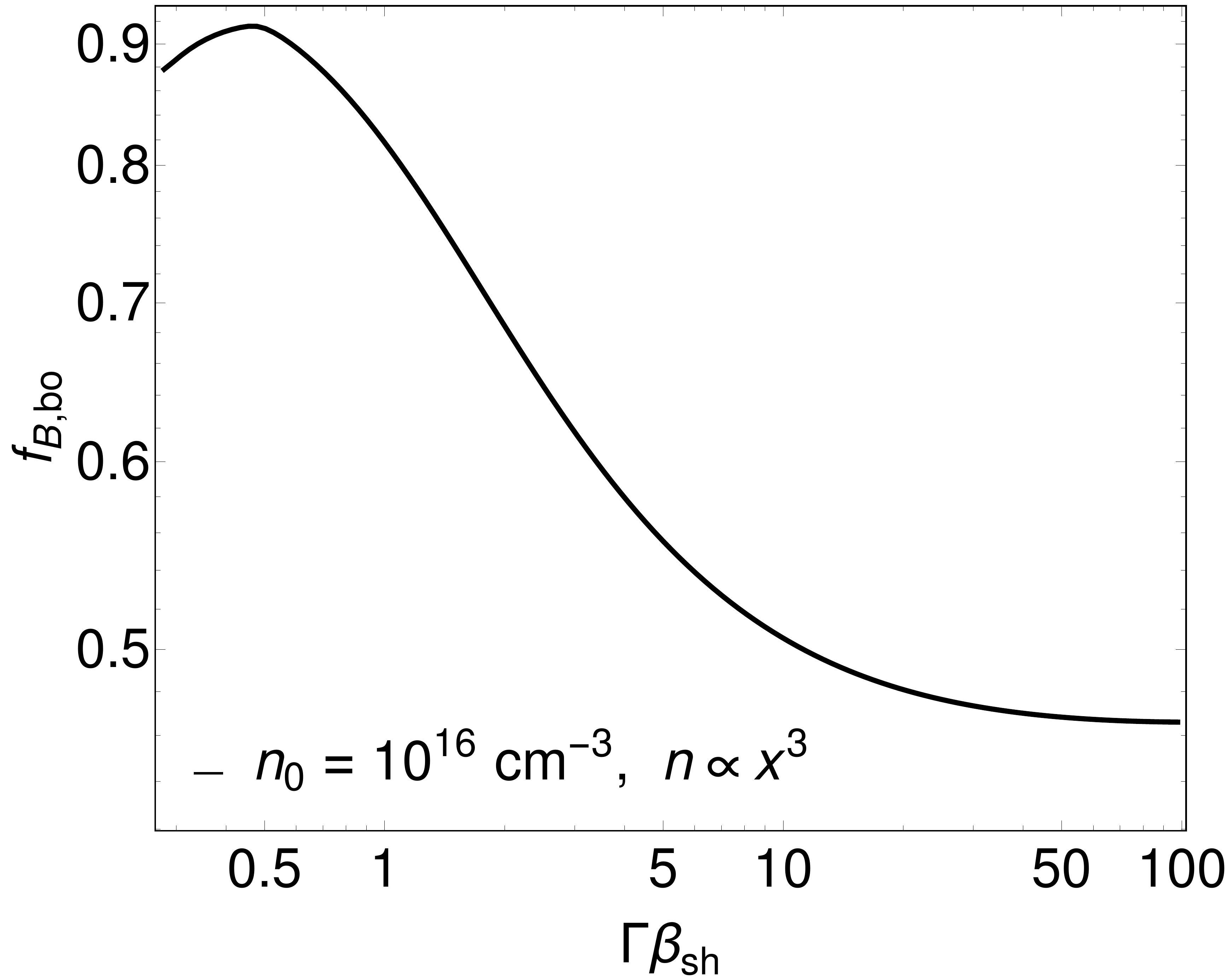} 
\caption{The fraction of soft free-free photons in the breakout shell that are upscattered to the Wien peak, computed for a case of a decreasing density profile with $n=3$ and $n_0 = 10^{16}$ cm $^{-3}$. At high shock velocities, the radiation has less time to Comptonize. The relatively high values of $f_\B$ indicate that the radiation in the breakout shell is very close to achieving a Wien spectrum, especially at moderate values of the shock Lorentz factor.}\label{f:fB_bo}
\end{figure*}

In a work in preparation, we compute the fraction of soft photons produces by free-free and double Compton emission that can be upscattered to the Wien peak in the breakout shell, denoted $f_\B$ and $f_\DC$, respectively. The breakout shell satisfies $\tau_\bo = 1/\beta_\d \simeq 3$, where $\beta_\d$ is the velocity of the flow in the immediate downstream in the shock frame. For that, we solve the following equation, in tandem with the equations described in Section \ref{sec:equations}:
    \begin{equation}\label{eq:ngammaW_eq}
        n_\gamma^\W = (f_\B\dot{n}^\ff+f_\DC\dot{n}^\DC)t\,,
    \end{equation}
where $f_\B$ and $f_\DC$ are functions of $\tau_\tot$, $\theta$ and $x_\m$ (see S84):
\begin{equation}
    f_\B = 2\Big[y_1^2-(y_1+y_1^2)\exp{(-1/y_1)}\Big]
\end{equation}
and
\begin{equation}
    f_\DC = y_1\big[1-\exp{(-1/y_1)}\big] ~,
\end{equation}
where
\begin{equation}
    y_1 = \frac{\tau_\tot^2(1+4\theta +16\theta^2)}{\ln \theta /x_\m}~
\end{equation}
is the modified Compton $y$ parameter, which marks the transition between saturated Comptonization ($y_1\gg 1$, $f_\B, f_\DC\sim 1$), for which a Wien distribution is obtained, and moderate Comptonization ($y_1<1$, $f_\B, f_\DC\ll 1$), where only a moderate fraction of the photons upscatter to $x=3\theta$. When $y_1\gg 1$, the spectrum receives the form of Wien spectral density per unit frequency $x$:
\begin{equation}\label{eq:Wien_spectrum}
    n_{\gamma,x}^\W = \frac{1}{2}n_\gamma\theta^{-1} \Big(\frac{x}{\theta}\Big)^2 \e^{-x/\theta} \,.
\end{equation}
In Figure \ref{f:fB_bo}, we show the solution for $f_\B$ as a function of the shock Lorentz factor at the breakout shell, where free-free emission dominates photon production. Already at the breakout shell the radiation achieves a spectrum that is very close to Wien. At higher optical depths, which are of interest in this work, photons are even more Comptonized due to the strong dependence of $f_\B$ on $\tau_\tot$, and therefore the assumption of a Wien spectrum applied in this work in deeper regions of the envelope is justified.

\section{Relativistic spherical hydrodynamics} \label{app:spherical_dynamics}
In this section we obtain the hydrodynamic behaviour of a fluid element experiencing spherical expansion, and find the terminal Lorentz factor that it can accelerate to given its initial Lorentz factor.
\subsection{Are shells expanding?}
We first check whether fluid elements change their width during spherical acceleration, or whether they keep an effectively constant width. The former is true if the Lorentz factor evolves slower than $\gamma\propto t'^{0.5}$.
The lab frame width and number density of a fluid element in the two cases are
\begin{equation}\label{eq:delta}
   \Delta'(t) = \Delta'_\s+\frac{c t'}{\gamma^2} \approx  \begin{cases}  \frac{c t'}{\gamma^2}, & \Delta'_\s\ll\frac{c t'}{\gamma^2} \\
   \Delta'_\s, & \frac{c t'}{\gamma^2} \ll \Delta'_\s\,,
	\end{cases}
\end{equation}
and
\begin{equation}
    n' \propto \frac{1}{r^2 \Delta'} = \frac{1}{(c t')^2\Delta'} \approx  \begin{cases} \frac{\gamma^2}{(ct')^3}, & \Delta'_\s\ll\frac{c t'}{\gamma^2}\\
    \frac{1}{(c t')^2\Delta'_\s}, & \frac{c t'}{\gamma^2} \ll \Delta'_\s
	\end{cases}
\end{equation}
respectively, where $\Delta'_\s$ is the initial width of a shell at the beginning of the spherical phase.
The time evolution of the fluid's number density, Lorentz factor and the pressure in the rest frame is written using the unknown power laws $\eta_\gamma,\, \eta_\n$  and $\eta_\p$:
\begin{subequations}
\begin{alignat}{3}
\gamma\propto t'^{\eta_\gamma}\label{eq:app_gamma_t}\\
 n\propto t'^{\eta_\n}\\
 P \propto n^{\gamma_\ad} \propto t'^{\eta_\p}
\end{alignat}
\end{subequations}
where $\eta_\p = \gamma_\ad\,\eta_\n$ and
\begin{equation}\label{eq:eta_p}
    \eta_\n = \begin{cases} \eta_\gamma-3, & \Delta'_\s\ll\frac{c t'}{\gamma^2}\\
    -2-\eta_\gamma, & \frac{c t'}{\gamma^2} \ll \Delta'_\s\,,
	\end{cases}
\end{equation}
where $\gamma_\ad$ is the adiabatic index of the fluid.
The generalized Riemann invariant (Oren \& Sari 2009) states that $    \gamma^2 \epsilon^{\pm \frac{\sqrt{3}}{2}} t^{\frac{4}{1+\sqrt{3}}} = const$. Therefore, we have another relation between $\eta_\p$ and $\eta_\gamma$:
\begin{equation}\label{eq:Riemann_gen}
    \eta_\gamma = -\frac{\sqrt{3}}{4}\eta_\p-\frac{2}{1+\sqrt{3}}~.
\end{equation}

\subsubsection{Option I: $\Delta' = \frac{c t'}{\gamma^2}$}
We now examine the case in which fluid elements significantly increase their width while accelerating. From equations \eqref{eq:eta_p} and \eqref{eq:Riemann_gen}, we have
\begin{equation}\label{eq:eta_gamma}
    \eta_\gamma = \frac{-8+9\gamma_\ad+3\sqrt{3}\gamma_\ad}{(1+\sqrt{3})(4+\sqrt{3}\gamma_\ad)} = \frac{\sqrt{3}}{1+\sqrt{3}}\,,
\end{equation}
where in the second equality we substituted $\gamma_\ad = 4/3$.
Using equations \eqref{eq:delta}, \eqref{eq:app_gamma_t} and \eqref{eq:eta_gamma} we find that
\begin{equation}
    \Delta' = \frac{c t'}{\gamma^2}\propto t'^{-2+\sqrt{3}} \rightarrow 0\,,
\end{equation}
which is in contradiction with our first assumption that $\Delta'_\s \ll \frac{ct'}{\gamma^2}$.

\subsubsection{Option II: $\Delta' = \Delta'_\s$}
In the same way as before, we have:
\begin{equation}
    \eta_\gamma = \frac{-2(-4+3\gamma_\ad+\sqrt{3}\gamma_\ad)}{(1+\sqrt{3})(-4+\sqrt{3}\gamma_\ad)}=1\,,
\end{equation}
where again we substituted $\gamma_\ad = 4/3$ in the second equality.
Checking for consistency with our initial assumption:
\begin{equation}
    \Delta' = \Delta'_\s+\frac{c t'}{\gamma^2} = \Delta'_\s+\mathcal{C}t'^{-1} \rightarrow \Delta'_\s \,,
\end{equation}
where $\mathcal{C}$ is a constant. We thus conclude that during the spherical phase, shells maintain their initial width, and the following holds: $\eta_\gamma = 1,\, \eta_p = -4,\, \eta_\n = -3$. We note that the same result was obtained by \cite{Piran1999}.
\subsection{Spherical acceleration and the terminal Lorentz factor}
Here we find the final Lorentz factor a fluid element obtains after it exhausts its internal energy. We define the following parameter $\mu \equiv \frac{\epsilon}{n m_\p c^2}$, which satisfies $\mu\gg1$ as long as the fluid is hot. The relations between $n$ and $\mu$ and between $\epsilon$ and $\mu$ for $\gamma_\ad = 4/3$ are
\begin{equation}
    n = n_\i\bigg(\frac{\mu}{\mu_\i}\bigg)^3
\end{equation}
and
\begin{equation}
    \epsilon = m_\p c^2 \mu n = m_\p c^2 n_\i \mu_\i \bigg(\frac{\mu}{\mu_\i}\bigg)^4 = \epsilon_\i\bigg(\frac{\mu}{\mu_\i}\bigg)^4 ~,
\end{equation}
respectively, where $\epsilon_\i = \gamma_\i m_\p c^2 n_\i$, and $\mu_\i = \gamma_\i$.
We use the planar phase Riemann invariant \citep{Johnson1971} to write $\epsilon$ and $\gamma$ at $t_\s$ in terms of the initial post shock properties. The planar phase Riemann invariant states that
\begin{equation}
    \frac{\sqrt{3}}{4}\log(\epsilon) \pm \frac{1}{2}\log(4\gamma^2) = const ~,
\end{equation}
giving
\begin{equation}
   \epsilon^{\sqrt{3}/2}_\i \gamma^2_\i =    \epsilon^{\sqrt{3}/2}_\s \gamma^2_\s ~,
\end{equation}
and
\begin{equation}\label{eq:mu_s}
    \mu_\s = \gamma_\i\bigg(\frac{\gamma_\i}{\gamma_\s}\bigg)^{1/\sqrt{3}} ~.
\end{equation}
Now, using the generalized Riemann invariant, we can find $\gamma_\f$:
\begin{equation}
  \bigg(\frac{\gamma_\f}{\gamma_\s}\bigg)^2\bigg(\frac{\epsilon_\f}{\epsilon_\s}\bigg)^{\sqrt{3}/2}\bigg(\frac{t'_\f}{t'_\s}\bigg)^{\frac{4}{1+\sqrt{3}}}  = \bigg(\frac{\gamma_\f}{\gamma_\s}\bigg)^2 \bigg(\frac{\mu_\f}{\mu_\s}\bigg)^{2\sqrt{3}}\bigg(\frac{t'_\f}{t'_\s}\bigg)^{\frac{4}{1+\sqrt{3}}} = 1 ~.
\end{equation}
Applying $\mu_\f = 1$ at the end of the acceleration phase, and using Eq \eqref{eq:mu_s}:
\begin{equation}
    \bigg(\frac{\gamma_\f}{\gamma_\s}\bigg)^2 \Bigg[\frac{1}{\gamma_\i\big(\frac{\gamma_\i}{\gamma_\s}\big)^{1/\sqrt{3}}}\Bigg]^{2\sqrt{3}}\bigg(\frac{t'_\f}{t'_\s}\bigg)^{\frac{4}{1+\sqrt{3}}}= 1~.
\end{equation}
Rearranging, we find:
\begin{equation}\label{eq:gammaf_tf}
    \gamma_\f = \gamma_\i^{1+\sqrt{3}}\bigg(\frac{t'_\f}{t'_\s}\bigg)^{-\frac{2}{1+\sqrt{3}}}  ~.
\end{equation}
The ratio $t'_\f/t'_\s$ can be found from
\begin{equation}
    \mu_\f = \mu_\s \bigg(\frac{n_\f}{n_\s}\bigg)^{1/3} = \mu_\s\bigg(\frac{t'_\f}{t'_\s}\bigg)^{-1} = 1\,.
\end{equation}
\begin{equation}
    \frac{t'_\f}{t'_\s} = \mu_\s  = \gamma_\i \bigg(\frac{\gamma_\i}{\gamma_\s}\bigg)^{1/\sqrt{3}}
\end{equation}
Substituting back into Eq \eqref{eq:gammaf_tf}:
\begin{equation}
    \gamma_\f = \gamma_\i^{1+\sqrt{3}} \Bigg[\gamma_\i \bigg(\frac{\gamma_\i}{\gamma_\s}\bigg)\Bigg]^{-\frac{2}{1+\sqrt{3}}} = \gamma_\i^{1+\frac{1}{\sqrt{3}}}\gamma_\s^{1-\frac{1}{\sqrt{3}}}~.
\end{equation}
Now, $\gamma_\s$ can be found using planar phase dynamics:
\begin{equation}
   t'_\i = \frac{\Delta'_\i}{c\Delta \beta} = \frac{\Delta'_\i}{c/\gamma^2}= \frac{x_\i/4\gamma^2}{c/\gamma^2} = \frac{x_\i}{4c} \propto \gamma_\i^{-2/\xi}
\end{equation}
where $x_\i$ is the width of a fluid element before it was shocked, the shell expands at a velocity $\Delta\beta \sim c/\gamma^2$ and $\xi$ is defined by $\Gamma_\sh \propto t'^{-\xi/2}$. Therefore,
\begin{equation}
    \gamma_\s = \gamma_\i \bigg(\frac{t'_\s}{t'_\i}\bigg)^{\frac{\sqrt{3}-1}{2}} = \gamma_\i\bigg(\frac{R/c}{x_\i/c}\bigg)^{\frac{\sqrt{3}-1}{2}} \propto \gamma_\i^{1+\frac{\sqrt{3}-1}{\xi}}\,.
\end{equation}
For $\xi = (2\sqrt{3}-3)n$ \citep{Sari2006}, we obtain:

\begin{equation}
\gamma_\f = \gamma_{\i,*}^{1+\sqrt{3}}\bigg(\frac{\gamma_\i}{\gamma_{\i,*}}\bigg)^{2+\frac{2}{3n}} ,
\end{equation}
where $\gamma_{\i,*}$ is the initial Lorentz factor of the shell that reaches its terminal Lorentz factor at $t_\s'$, and for $n=3$ it is equal to
\begin{equation}\label{eq:gs_gf}
    \gamma_{\i,*} = 1.6\, (n_0 \sigma_\T  R_*)^{0.30} \gamma_{\i,0}^{-0.44}\,.
\end{equation}
After elements have converted all of their internal energy to bulk acceleration, they enter the cold spherical phase. During this phase, the Lorentz factor is constant, and therefore $\eta_\gamma = 0$.
This implies that $\Delta' \propto t'$ and shells increase their width. The time evolution is the rest frame remains the same as in the hot phase.

\section{Work exerted on an electron by the radiation field}\label{sec:decoupling}
The work done on an electron by a radiation field, assumed to move in a specific direction, is the heating rate multiplied by the interaction time:
\begin{equation}
\begin{split}
    W &= \sigma_\T \Delta v \epsilon'\frac{r}{c} = \sigma_\T \frac{c}{\gamma^2}(\gamma^2\epsilon)\frac{r}{c} =\sigma_\T \epsilon r = \sigma_\T \Bigg[\gamma^2_\i4 m_\p c^2 n_\i\bigg(\frac{t_\s}{t_\i}\bigg)^{-4/3}\bigg(\frac{t'}{t'_\s}\bigg)^{-4}\Bigg]r \\
    &=\sigma_\T (ct'_\s) \bigg(\frac{t_\s}{t_\i}\bigg)^{-4/3}\bigg(\frac{t'}{t'_\s}\bigg)^{-3}\Big(4\gamma^2_\i n_\i m_\p c^2\Big) = \sigma_\T (ct'_\s) \bigg(\frac{\gamma_\s}{\gamma_\i}\bigg)^{-4/\sqrt{3}}\bigg(\frac{\gamma}{\gamma_\s}\bigg)^{-3}\Big(4\gamma^2_\i n_\i m_\p c^2\Big)~.
\end{split}
\end{equation}
In order for the electrons to be accelerated to the Lorentz factor of the local radiation field, $\gamma$, we must require that $W\geq \gamma m_\e c^2$, which translates to
\begin{equation}
\gamma \leq \bigg(4 \frac{m_\p}{m_\e} \sigma_\T n_\i R_*\bigg)^{1/4} \gamma_\i^{\frac{1}{2}+\frac{1}{\sqrt{3}}}\gamma_\s^{\frac{3}{4}-\frac{1}{\sqrt{3}}
}~.
\end{equation}
The above expression can be translated to the mass of the shell beyond which the fluid can no longer be accelerated to the Lorentz factor of the radiation field, which we denote as the `decoupling shell':
\begin{equation}
m_\dc \simeq 10^{-3}\,m_0 \bigg(\frac{R_*}{5 \text{R}_\odot}\bigg)^{0.28}\bigg(\frac{M_\ej}{5 \text{M}_\odot}\bigg)^{0.56} \bigg(\frac{E_\exp}{10^{53} \text{erg}}\bigg)^{-0.71}  \bigg(\frac{t'}{t'_\s}\bigg)^{4.55} ~.
\end{equation}
If at some point $m_\tr<m_\dc$, there will be two characteristics Lorentz factors: one of the radiation field and anther of the matter. Since during the spherical phase $m_\tr = m_0 \Big(\frac{t'}{t'_\s}\Big)^2$, the luminosity shell is initially more massive than the decoupling shell, and the two shells intersect at
\begin{equation}
    t'_\dc = 16\, t'_\s\, \bigg(\frac{R_*}{5 \text{R}_\odot}\bigg)^{-0.11}\bigg(\frac{M_\ej}{5 \text{M}_\odot}\bigg)^{-0.22} \bigg(\frac{E_\exp}{10^{53} \text{erg}}\bigg)^{0.28}  ~.
\end{equation}
In principle, at $t'_\dc<t'$ the luminosity shell is shallower than the decoupling shell. However, the fluid typically cools before this time is reached (see Eq \ref{eq:tc_sph_lab}), and the Lorenz factor remains constant. Therefore, fluid elements reach their terminal Lorentz factor long before they decouple from radiation.

\section{Glossary}
General notations for any quantity $q$:
\begin{itemize}[label={}]
  \item $q$: a quantity measured in the fluid rest frame.
  \item $q'$: a quantity measured in the lab frame.
    \item $q^\obs, q_\obs$: a quantity measured in the observer frame, if different from the lab frame.
  \item $q_\i$: a quantity measured in the immediate downstream of the shock.
  \item $q_\tr$: a property of the luminosity shell.
  \item $q_0$: a property of the unloaded breakout shell, that has a pre-explosion pair-unloaded optical depth of unity.
  \item $q_\sh$: a property of the shock front.
  \item $q_\c$: a property of the first shell that was exposed cold during the planar phase.
\end{itemize}
General physical variables:
\begin{itemize}[label={}]
  \item $t$: rest frame time since shock breakout.
  \item $t'$: lab frame time since shock breakout.
  \item $v$: velocity.
  \item $\beta$: velocity in units of the speed of light.
\item $\Gamma, \gamma$: the Lorentz factor of the shock and fluid, respectively.
\item $n$: number density.
\item $\rho$: mass density.
\item $\epsilon$: energy density.
\item $L$: luminosity.
\item $T$: temperature.
\item $\theta$: temperature in units of the electron rest energy.
\item $\tau$: optical depth.
\item $d$: shell width, also the distance to the edge of the stellar envelope.
\item $\sigma_\T$: Thomson scattering cross section.
\item $\nu$: frequency.
\item $x$: photon energy in units of the electron rest energy.
\end{itemize}

\begin{itemize}[label={}]
\item $M_\ej$: ejecta mass.
\item $R_*$: the stellar radius prior to shock breakout.
\end{itemize}

Specific notations and symbols:
\begin{itemize}[label={}]
  \item $n_-$: electron number density. 
  \item $n_+$: positron number density. 
  \item $n_\p$: proton number density.
  \item $z_+$: the positron fraction relative to protons.
  \item $\tau_\T$: Thomson optical depth.
  \item $x_\m$: the photon energy below which the photon spectrum is a Planckian, in units of the electron rest energy.
  \item $\tau_\tot$: the pair-loaded electron scattering optical depth.
  \item $m_{\c,\tr}$: the mass of the first shell that is exposed cold during the planar phase.
  \item $t_\i$: the initial dynamical time of a shell after shock breakout.
  \item $t_\s,t'_\s,t_\s^\obs$: the transition time to the spherical phase in the rest frame, the lab frame, and the observed frame, respectively.
  \item $t_\c^\obs$: the observed time after which the escaping planar phase radiation originates from cold shells.
  \item $t_{\c,2}^\obs$: the observed time at which the first shell that was exposed cold during the planar phase stops being the main luminosity source.
  \item $t_{\c,\s}^\obs$: the observed time after which the escaping spherical phase radiation originates from cold shells.
  \item $t_{\c,\pl}^\obs$: the observed time after which spherical phase radiation originates in shells that have cooled during the planar phase.
  \item $\gamma_\f$: the final Lorentz factor of a shell after it has exhausted its internal energy.
  \item $t_{z_+,\i}^\obs$: the observed time after which exposed shells were never pair loaded.
  \item $t_\NW^\obs$: the observed time when the dynamics of exposed shells become Newtonian.
\end{itemize}

\bibliography{relativistic_breakout}{}
\bibliographystyle{aasjournal}
\end{document}